\documentclass{amsart}
\usepackage{amsfonts,amsmath,amssymb}
\pdfoutput=1
\usepackage[longnamesfirst,round]{natbib}
\usepackage{multirow,graphicx}
\usepackage{array,rotating}

\newcommand{\e}{\mathbb{E}}

\newcommand{\bs}{\boldsymbol}
\newcommand{\mbf}{\mathbf}
\newcommand{\diag}{\mathop{\mathrm{diag}}}

\newcommand{\rank}{\mathop{\mathrm{rank}}}
\newcommand{\spn}{\mathop{\mathrm{span}}}
\newcommand{\nrm}{\mathcal{N}}

\newcommand{\ber}{\mathcal{B}}
\newcommand{\poi}{\mathcal{P}}

\title[Dimension Reduction and Alleviation of Confounding for SGLMMs]{Dimension Reduction and Alleviation of Confounding for Spatial Generalized Linear Mixed Models}  
\author[Hughes and Haran]{John Hughes and Murali Haran\\The Pennsylvania State University, USA}

\begin{document}

\begin{abstract}
Non-gaussian spatial data are very common in many disciplines. For instance, count data are common in disease mapping, and binary data are common in ecology. When fitting spatial regressions for such data, one needs to account for dependence to ensure reliable inference for the regression coefficients. The spatial generalized linear mixed model (SGLMM) offers a very popular and flexible approach to modeling such data, but the SGLMM suffers from three major shortcomings: (1) uninterpretability of parameters due to spatial confounding, (2) variance inflation due to spatial confounding, and (3) high-dimensional spatial random effects that make fully Bayesian inference for such models computationally challenging. We propose a new parameterization of the SGLMM that alleviates spatial confounding and speeds computation by greatly reducing the dimension of the spatial random effects. We illustrate the application of our approach to simulated binary, count, and Gaussian spatial datasets, and to a large infant mortality dataset. 
\end{abstract}

\maketitle


\section{Introduction} 
\label{intro}

Nearly 20 years ago \citet{Besag:1991p873} proposed the spatial generalized linear mixed model (SGLMM) for areal data, a hierarchical model that introduces spatial dependence through a latent Gaussian Markov random field (GMRF). Although \citeauthor{Besag:1991p873} focused only on prediction for count data, the SGLMM for areal data has since been applied to other types of data (e.g., binary), in many fields (e.g., ecology, geology, forestry), and for regression as well as prediction. In fact, the SGLMM has long been the dominant areal model, owing to its flexible hierarchical specification; to the availability of the WinBUGS software application \citep{Lunn:2000p884}, which greatly eases data analysis; and to the various theoretical and computational difficulties that plague the areal SGLMM's nearest competitor, the automodel \citep[see, e.g.,][]{Hughes2011Autologistic-Mo,Kaiser:1997p812,Kaiser:2000p352}.

SGLMMs, while remarkably flexible and widely applicable, suffer from three major shortcomings: (1) uninterpretability of parameters due to spatial confounding, (2) variance inflation due to spatial confounding, and (3) computational challenges posed by high-dimensional latent variables. The issue of spatial confounding between the fixed and random effects was pointed out recently by \citet{BrianJReich:2006p787}, and they suggested a new model (henceforth RHZ) that seeks to alleviate confounding by explicitly introducing synthetic predictors that are orthogonal to the fixed effects predictors. But the RHZ model, by failing to account for the underlying graph, permits structure in the random effects that corresponds to negative spatial dependence, i.e., repulsion, which we do not expect to observe in the phenomena to which these models are typically applied.

We propose a new model that mitigates confounding while allowing only positive spatial dependence among the random effects. We achieve this by exploiting what we believe to be the ``natural" geometry for these models. By utilizing this geometry we are also able to dramatically reduce the dimension of the random effects, which speeds computation to the extent that our model makes feasible the analyses of large areal datasets.

Several recent papers have focused on dimension reduction for point-level, i.e., Gaussian process-based, spatial models \citep[cf., e.g.,][]{Higdon:2002p909,Cressie:2008p911,Banerjee:2008p903,Furrer:2006p912,Rue:2002p915}. To our knowledge, this paper is the first to propose a principled approach to dimension reduction for areal models (while also improving regression inference).

The rest of the paper is organized as follows. In Section~\ref{traditional} we review the traditional SGLMM. In Section~\ref{confounding} we discuss spatial confounding and review the model proposed by \citeauthor{BrianJReich:2006p787} In Section~\ref{sparse} we present our sparse reparameterization of the areal SGLMM. In Section~\ref{reduction} we discuss dimension reduction for spatial models. In Section~\ref{simulation} we discuss the results of our simulation study, which applied each of the three models to simulated binary, count, and Gaussian spatial data. In Section~\ref{real} we apply our sparse model to a large US infant mortality dataset. We conclude with a discussion.

\section{The Traditional SGLMM}
\label{traditional}

\citet{Besag:1991p873} formulated their SGLMM for areal data as follows. Let $G=(V,E)$ be the underlying undirected graph, where $V=\{1,2,\ldots,n\}$ are the vertices (each of which represents an area of interest) and $E$ are the edges (each of which is a pair, $(i,j)$, that represents the proximity of areas $i$ and $j$). In the sequel we shall represent $G$ using its adjacency matrix, $\mbf{A}$, which is the $n\times n$ matrix with entries given by $\diag(\mbf{A})=\bs{0}$ and $\mbf{A}_{ij}=1\{(i,j)\in E,i\neq j\}$, where $1\{\cdot\}$ denotes the indicator function.

Now, let $\bs{Z}=(Z_1,\ldots,Z_n)^\prime$ be the random field of interest, where $Z_i$ is the random variable associated with vertex $i$. Then the first stage of the model is given by
\begin{align}
\label{tradfs}
g(\e(Z_i\,|\,\bs{\beta},W_i)) &= \mbf{X}_i\bs{\beta}+W_i,
\end{align}
where $g$ is a link function, $\mbf{X}_i$ is the $i$th row of the design matrix, $\bs{\beta}$ is a $p$-vector of regression parameters, and $W_i$ is a spatial random effect associated with vertex $i$. As is the case for classical generalized linear models, different types of data imply different (canonical) choices of the link function, $g$. For example, the canonical link function for binary spatial data is the logit function, $\text{logit}(p)=\log(p/(1-p))$, for count data it is the natural logarithm function, and for normal data it is the identity function \citep{Nelder:1972p908}. In the latter case we have the spatial linear mixed model (SLMM), which is of considerable interest \citep[see, e.g.,][]{cres:1993,ban:carl:gelf:2004}.

The field of random effects, $\bs{W}=(W_1,\ldots,W_n)^\prime$, whereby the traditional model incorporates spatial dependence, is assumed to follow the so-called conditionally autoregressive (CAR) or Gaussian Markov random field prior:
\begin{align}
\label{car}
p(\bs{W}\,|\,\tau) &\propto \tau^{\rank(\mbf{Q})/2}\exp\left(-\frac{\tau}{2}\bs{W}^\prime\mbf{Q}\bs{W}\right),
\end{align}
where $\tau$ is a smoothing parameter and $\mbf{Q}=\diag(\mbf{A}\bs{1})-\mbf{A}$ is a precision matrix ($\bs{1}$ is the conformable vector of 1s). Note that $\mbf{Q}$ intuitively incorporates both dependencies ($W_i$ and $W_j$, $i\neq j$, are independent given their neighbors iff $\mbf{Q}_{ij}=\mbf{Q}_{ji}=0$ iff $(i,j)\notin E$) and prior uncertainty (our uncertainty about $W_i$ is inversely proportional to the degree of vertex $i$: $\mbf{Q}_{ii}=\mbf{A}_i\bs{1}$).

As described in the seminal paper by \citet{Diggle:1998p264}, an SGLMM for spatial data over a continuous domain, i.e., for point-level data, can be formulated by replacing the GMRF specification with that of a Gaussian process \citep[also see][]{Oliveira:2000p266,Christensen:2002p259}. See \citet{ban:carl:gelf:2004}, \citet{Rue:Held:2005}, or \citet{hara:2009} for an introduction to the use of Gaussian random field models in spatial statistics.

Since (\ref{car}) is improper ($\mbf{Q}$ is singular) the traditional areal model restricts us to a Bayesian analysis, although it is possible to use a maximum likelihood-based approach if a proper GMRF distribution is used instead of this so-called intrinsic GMRF \citep[see][]{Besag:1995p891}. The models described below in Sections~\ref{confounding} and \ref{sparse} have CARs with invertible precision matrices, and so those models lend themselves to both  classical and Bayesian analyses. In the interest of concision we use only Bayesian terminology in the sequel.

\section{Spatial Confounding}
\label{confounding}

We begin with a discussion of confounding for the areal SGLMM's closest competitor, the automodel, because we believe this offers insight regarding confounding for the SGLMM. The automodel, a Markov random field model that incorporates dependence directly rather than hierarchically, can be specified as
\begin{align}
\label{auto}
g(\e(Z_i\,|\,\bs{\beta},\eta,\bs{Z}_{-i})) &= \mbf{X}_i\bs{\beta}+\eta\sum_{(i,j)\in E}Z_j^*,
\end{align}
where $\bs{Z}_{-i}$ denotes the field with the $i$th observation excluded and $\eta$ is the dependence parameter ($\eta>0$ implies an attractive model, $\eta<0$ a repulsive model). For the so-called uncentered automodel, $Z_j^*=Z_j$, while $Z_j^*=Z_j-\mu_j$ for the centered automodel, where $\mu_j$ is the independence expectation of $Z_j$, i.e., $\mu_j=\e(Z_j\,|\,\bs{\beta},\eta=0)=g^{-1}(\mbf{X}_j\bs{\beta})$.

The sum in (\ref{auto}) is called the autocovariate. We see that it is a synthetic predictor that employs only the observations themselves or the observations along with the posited regression component, $\mbf{X}\bs{\beta}$, of the model. The centered form of the autocovariate leads easily to an interpretation of the dependence parameter: $\eta$ measures the ``reactivity" of an observation to its neighbors, conditional on the hypothesized regression component. This interpretation of $\eta$ lends $\bs{\beta}$ their desired interpretation as regression parameters. Put another way, the purpose of the centered autocovariate is to fit small-scale structure in the data, by which we mean structure residual to the large-scale structure represented by $\mbf{X}\bs{\beta}$. Evidently the centered autocovariate is well suited to this role, since $\eta$ tends to be at most weakly correlated with $\bs{\beta}$. 

The uncentered automodel, on the other hand, exhibits both conceptual and spatial confounding. It is not clear how one should interpret the uncentered autocovariate, and so $\eta$ and $\bs{\beta}$ are also difficult to interpret. Moreover, $\eta$ and $\bs{\beta}$ tend to be strongly correlated for the uncentered model. (See \citet{Caragea:2009p778} for a treatment of confounding in the context of the autologistic model.)


\citet{BrianJReich:2006p787} employed a reparameterization to show that the traditional SGLMM, like the uncentered automodel, is confounded. More specifically, they showed that introduction of the random effects tends to bias the posterior distribution of $\bs{\beta}$ and also inflate its variance. This is because the traditional model implicitly contains predictors that are collinear with $\mbf{X}$, and it is this collinearity that causes the bias and variance inflation. To see this, let $\mbf{P}$ be the orthogonal projection onto $\spn(\mbf{X})$, i.e., $\mbf{P}=\mbf{X}(\mbf{X}^\prime\mbf{X})^{-1}\mbf{X}^\prime$, and let $\mbf{P}^\perp$ be the projection onto $\spn(\mbf{X})$'s orthogonal complement: $\mbf{P}^\perp=\mbf{I}-\mbf{P}$. Now, spectrally decompose these operators to acquire orthogonal bases, $\mbf{K}_{n\times p}$ and $\mbf{L}_{n\times (n-p)}$, for $\spn(\mbf{X})$ and $\spn(\mbf{X})^\perp$, respectively. These bases allow us to rewrite (\ref{tradfs}) as
\begin{align*}
g(\e(Z_i\,|\,\bs{\beta},W_i)) &= \mbf{X}_i\bs{\beta}+W_i = \mbf{X}_i\bs{\beta}+\mbf{K}_i\bs{\gamma}+\mbf{L}_i\bs{\delta},
\end{align*}
where $\bs{\gamma}$ and $\bs{\delta}$ are random coefficients. This form exposes the source of the spatial confounding: $\mbf{K}$ is collinear with $\mbf{X}$.

Since the offending predictors, $\mbf{K}$, have no scientific meaning, \citeauthor{BrianJReich:2006p787} suggest that they be deleted from the model. Setting $\bs{\gamma}=\bs{0}$ leads to the following specification. For the first stage we have
\begin{align*}
g(\e(Z_i\,|\,\bs{\beta},\bs{\delta})) &= \mbf{X}_i\bs{\beta}+\mbf{L}_i\bs{\delta}.
\end{align*}
And the prior for the random effects, $\bs{\delta}$, is now
\begin{align*}
p(\bs{\delta}\,|\,\tau) &\propto \tau^{(n-p)/2}\exp\left(-\frac{\tau}{2}\bs{\delta}^\prime\mbf{Q}_R\bs{\delta}\right),
\end{align*}
where $\mbf{Q}_R=\mbf{L}^\prime\mbf{Q}\mbf{L}$.

\citeauthor{BrianJReich:2006p787} refer to this as smoothing orthogonal to the fixed effects, and they showed that their model does, in principle, address the aforementioned problems. More specifically, adding this restricted CAR to the nonspatial model adjusts, but does not inflate, the variance of $\bs{\beta}$'s posterior but leaves the mean unchanged. We also note that RHZ reduces slightly the number of model parameters, from $n+p+1$ to $n+1$.

The traditional SGLMM and RHZ take essentially the same approach as the automodel in the sense that all of these models augment the linear predictor with synthetic predictors. The traditional model can be considered to implicitly employ the synthetic predictors $\mbf{K}$ and $\mbf{L}$, and RHZ explicitly employs $\mbf{L}$. The traditional SGLMM is analogous to the uncentered automodel in that both models introduce predictors---$\mbf{K}$ and the uncentered autocovariate, respectively---that cause spatial confounding and thus render $\bs{\beta}$ uninterpretable. And RHZ is analogous to the centered automodel in that both models introduce predictors designed to fit only residual structure in the data.

\section{A Sparse Reparameterization of the Areal SGLMM}
\label{sparse}

Although the synthetic predictors for RHZ are readily interpreted, receiving their interpretation from the theory of linear models, they are not particularly well suited to their task of fitting the residual clustering that arises due to spatial dependence. This is because the geometry corresponding to the operator $\mbf{P}^\perp$ neglects the underlying graph, $G$. In this section we reparameterize the areal SGLMM using an alternative operator that we believe corresponds to the intrinsic geometry of these models. The random effects for our new model (1) permit patterns corresponding to positive spatial dependence only, i.e., repulsion is disallowed, and (2) have dimension much smaller than $n$.

In an attempt to reveal the structure of missing spatial covariates, \citet{Griffith2003Spatial-Autocor} augments a generalized linear model with selected eigenvectors of $(\mbf{I}-\bs{1}\bs{1}^\prime/n)\mbf{A}(\mbf{I}-\bs{1}\bs{1}^\prime/n)$, where $\mbf{I}$ is the $n\times n$ identity matrix and $\bs{1}$ is the $n$-vector of 1s. This operator appears in the numerator of Moran's $I$ statistic,
\begin{align*}
I(\mbf{A}) &= \frac{n}{\bs{1}^\prime\mbf{A}\bs{1}}\frac{\bs{Z}^\prime(\mbf{I}-\bs{1}\bs{1}^\prime/n)\mbf{A}(\mbf{I}-\bs{1}\bs{1}^\prime/n)\bs{Z}}{\bs{Z}^\prime(\mbf{I}-\bs{1}\bs{1}^\prime/n)\bs{Z}},
\end{align*}
a popular nonparametric measure of spatial dependence \citep{Moran:1950p874}. We see that $I(\mbf{A})$ is a scaled ratio of quadratic forms in $\bs{e}=\bs{Z}-\bar{Z}\bs{1}$. The numerator of the ratio is the squared length of $\bs{e}$ in the elliptical space corresponding to $\mbf{A}$, while the denominator is the squared length of $\bs{e}$ in a spherical space.

Since we are interested not in missing covariates but in smoothing orthogonal to $\mbf{X}$ (which may or may not contain $\bs{1}$), we replace $\mbf{I}-\bs{1}\bs{1}^\prime/n$ with $\mbf{P}^\perp$. The resulting operator, $\mbf{P}^\perp\mbf{A}\mbf{P}^\perp$, which we call the Moran operator for $\mbf{X}$ with respect to $G$, appears in the numerator of a generalized form of Moran's $I$:
\begin{align*}
I_\mbf{X}(\mbf{A}) &= \frac{n}{\bs{1}^\prime\mbf{A}\bs{1}}\frac{\bs{Z}^\prime\mbf{P}^\perp\mbf{A}\mbf{P}^\perp\bs{Z}}{\bs{Z}^\prime\mbf{P}^\perp\bs{Z}}.
\end{align*}
\citet{Boots:2000p914} showed that (1) the (standardized) spectrum of a Moran operator comprises the possible values for the corresponding $I_\mbf{X}(\mbf{A})$, and (2) the eigenvectors comprise all possible mutually distinct patterns of clustering residual to $\mbf{X}$ and accounting for $G$. The positive (negative) eigenvalues correspond to varying degrees of positive (negative) spatial dependence, and the eigenvectors associated with a given eigenvalue ($\lambda_i$, say) are the patterns of spatial clustering that data exhibit when the dependence among them is of degree $\lambda_i$.

\begin{figure}
   \centering
   \begin{tabular}{cc}
   \includegraphics[scale=.6]{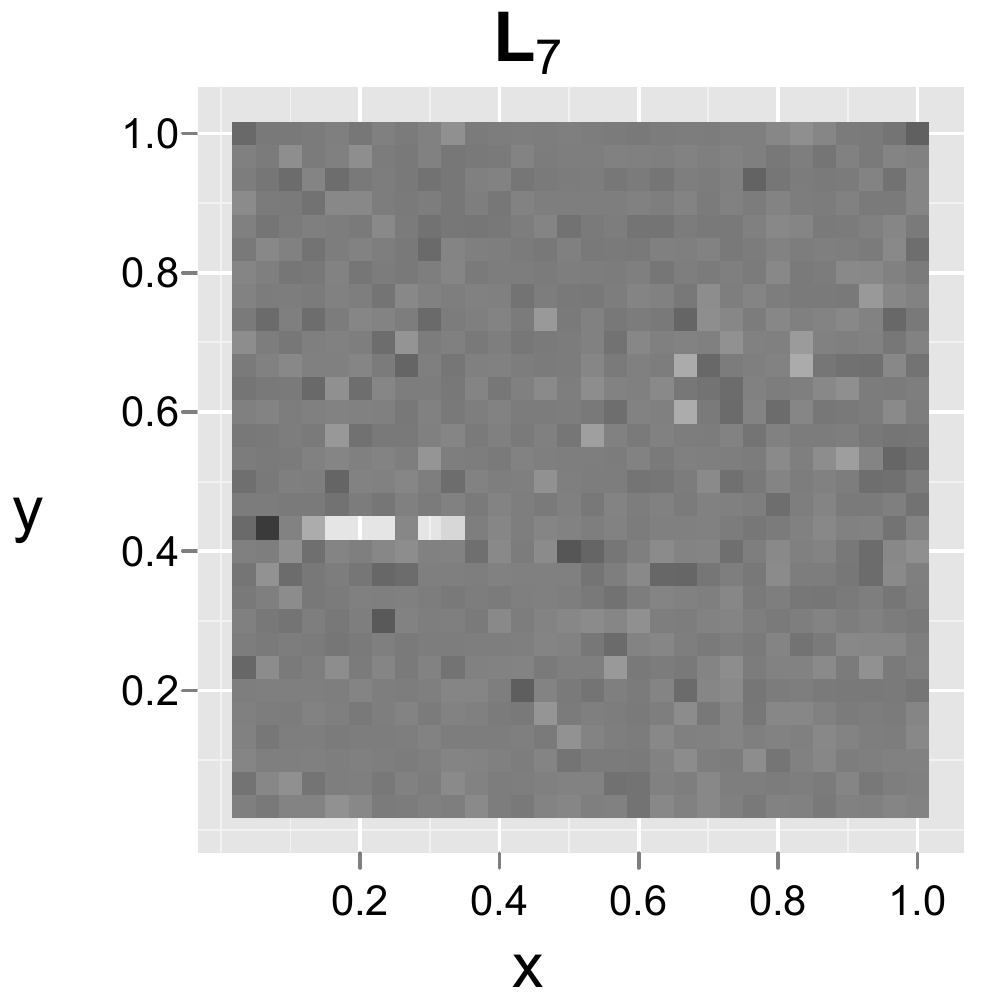} & \includegraphics[scale=.6]{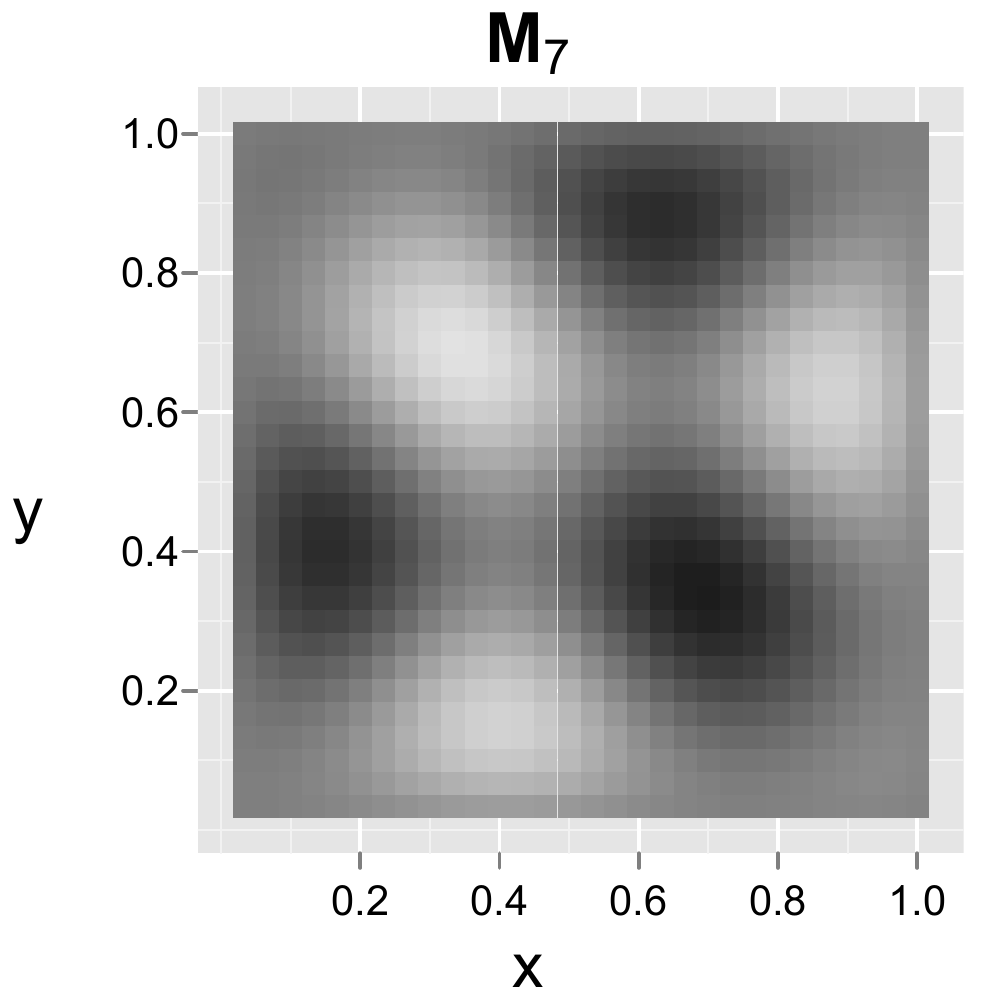} \\
   \includegraphics[scale=.6]{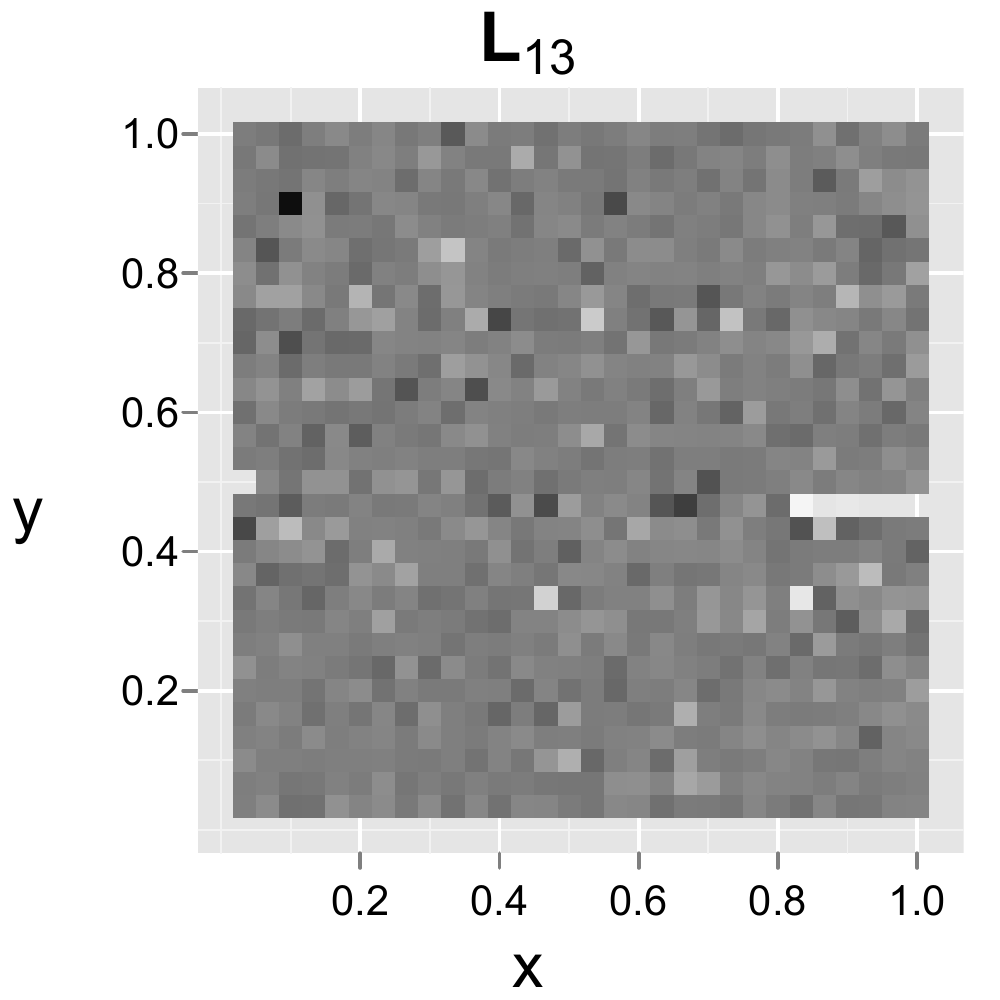} & \includegraphics[scale=.6]{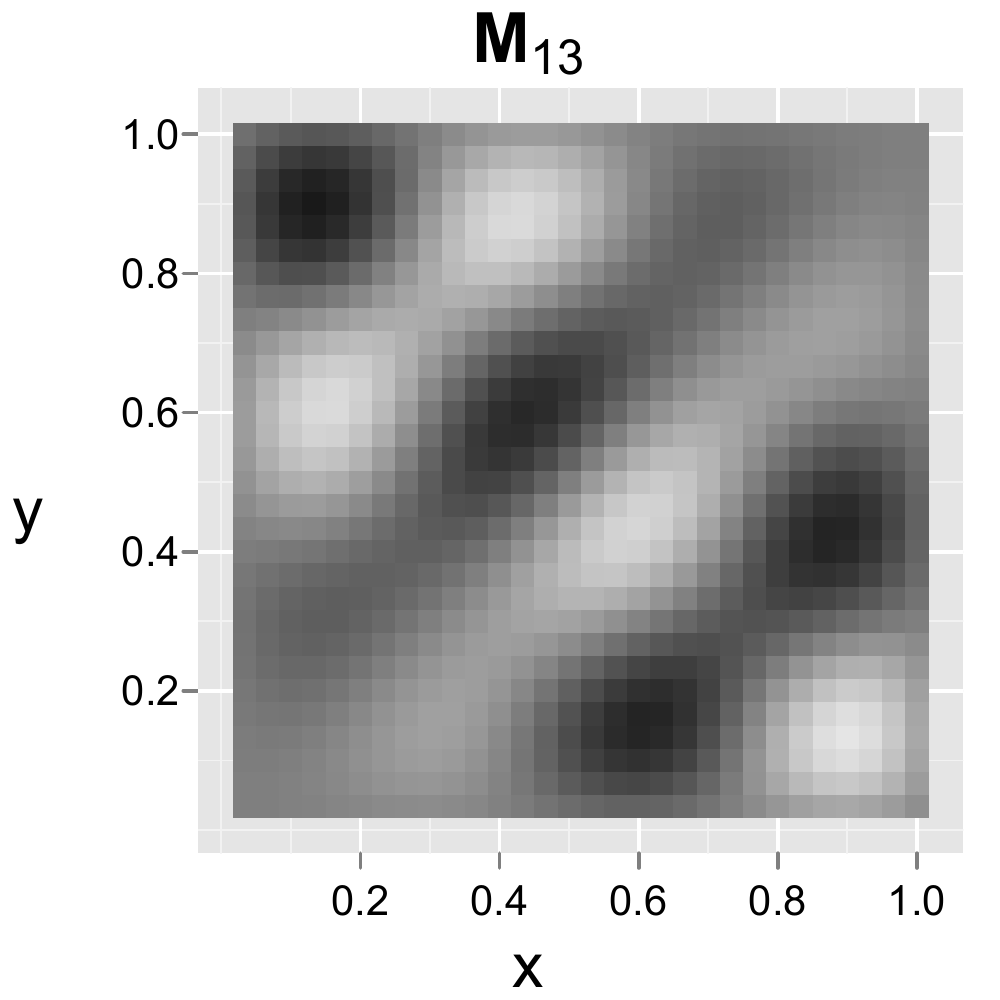} \\
   \includegraphics[scale=.6]{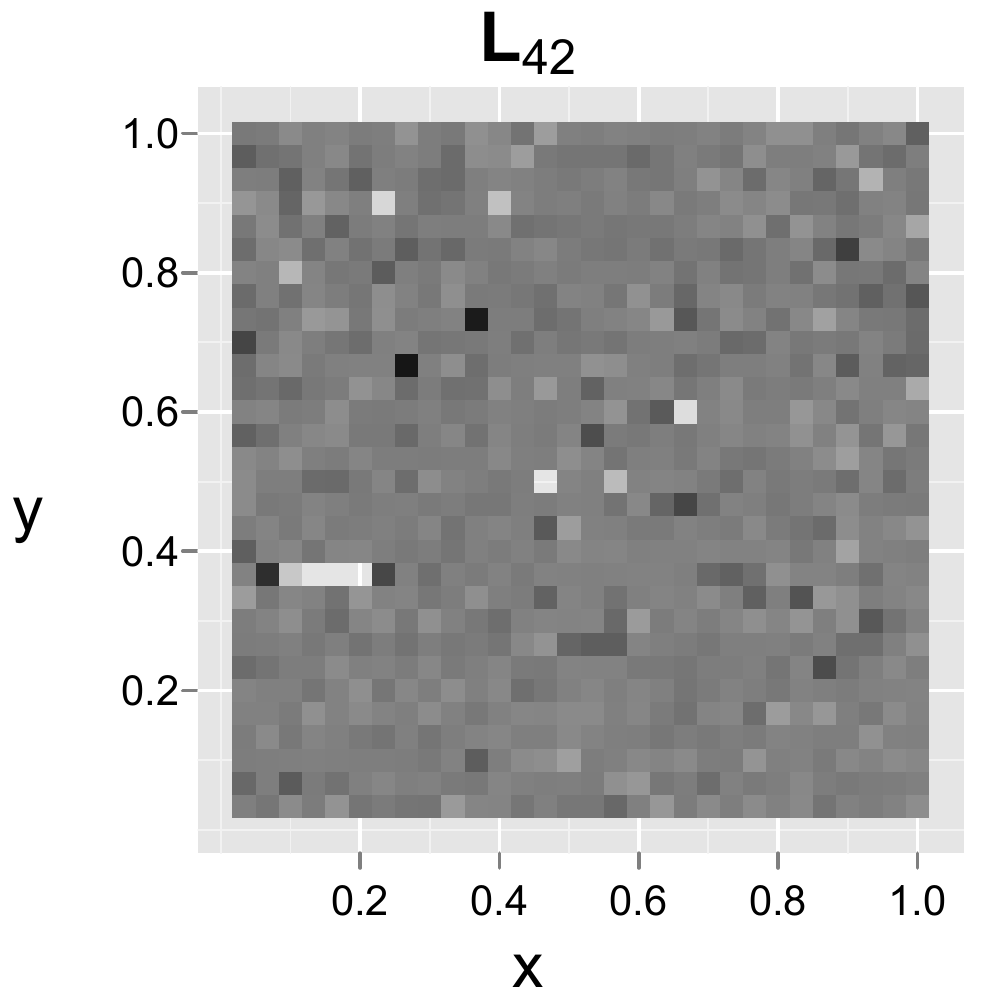} & \includegraphics[scale=.6]{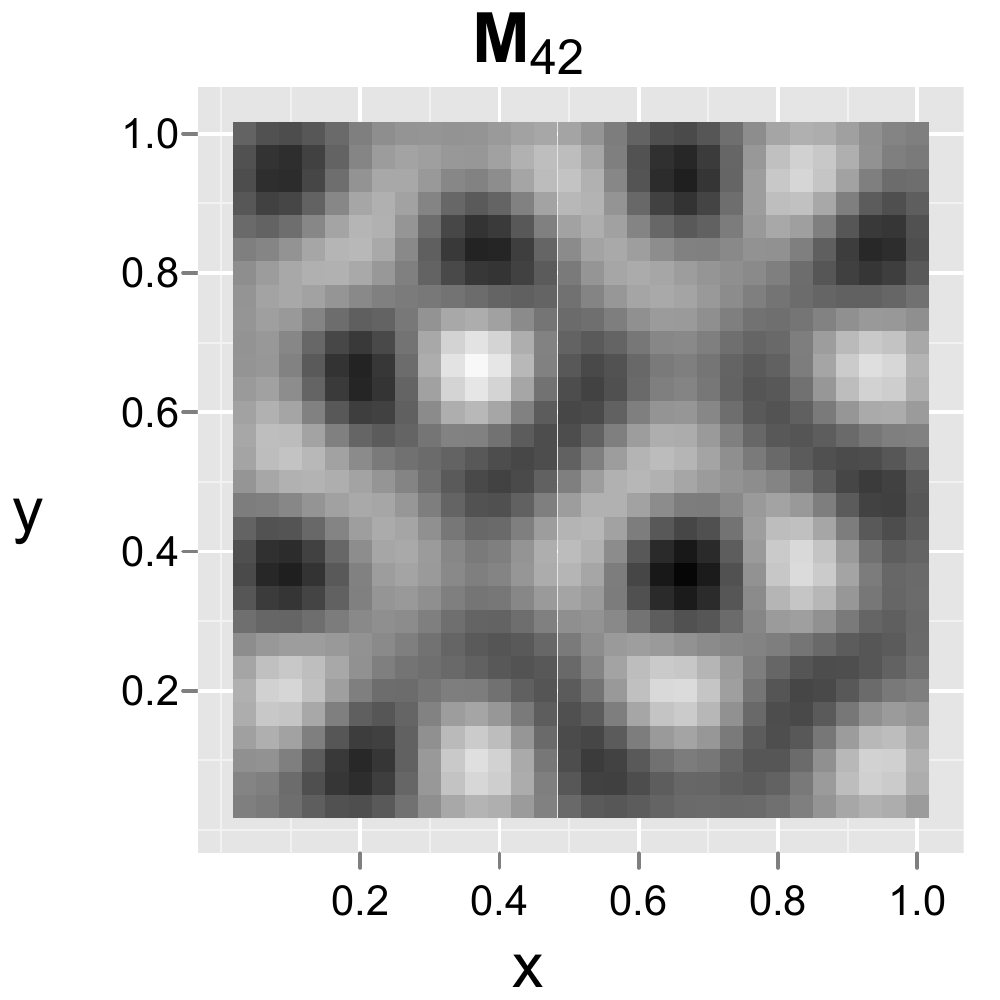} \\
   \end{tabular}
   \caption{\label{vectors}Eigenvectors 7, 13, and 42 from the RHZ and Moran bases. The Moran basis appears to be an appropriate basis for fitting small-scale structure.}
\end{figure}

To illustrate the appropriateness of the Moran basis for fitting residual spatial clustering, we consider the $30\times 30$ square lattice and $\mbf{X}=[\bs{x}\;\bs{y}]$, where the $i$th row of the design matrix contains the $x$ and $y$ coordinates of the $i$th lattice point, with coordinates restricted to the unit square. The panels of Figure~\ref{vectors} show eigenvectors 7, 13, and 42 (which we chose at random) of the RHZ basis ($\mbf{L}_7,\mbf{L}_{13},\mbf{L}_{42}$) and of the Moran basis ($\mbf{M}_7,\mbf{M}_{13},\mbf{M}_{42}$). Each panel displays its eigenvector as a ``map" by associating the $i$th component of the vector with the spatial location, $(x_i,y_i)$, of the $i$th lattice point. (We note that $\mbf{M}_7$, $\mbf{M}_{13}$, and $\mbf{M}_{42}$ are associated with eigenvalues 0.995, 0.970, and 0.868, respectively, which implies that these eigenvectors correspond to strong positive spatial dependence.) Each vector of the RHZ basis is relatively flat, with the flatness occasionally broken by a very localized spike or dip. The vectors of the Moran basis, on the other hand, show clear and regular patterns of spatial variation at various scales. 

We henceforth assume that the matrix $\mbf{M}$ contains $q\ll n$ eigenvectors of the Moran operator. Since eigenvectors that correspond to inconsequential positive dependence, independence, or negative dependence (repulsion) are of no interest, we can discard at least half of the eigenvectors in many cases, and our simulation study will show that a much greater reduction is often possible.

Replacing $\mbf{L}$ with $\mbf{M}$ in the RHZ model gives, for the first stage,
\begin{align*}
g(\e(Z_i\,|\,\bs{\beta},\bs{\delta}_S)) &= \mbf{X}_i\bs{\beta}+\mbf{M}_i\bs{\delta}_S.
\end{align*}
And the prior for the random effects is now
\begin{align*}
p(\bs{\delta}_S\,|\,\tau) &\propto \tau^{q/2}\exp\left(-\frac{\tau}{2}\bs{\delta}_S^\prime\mbf{Q}_S\bs{\delta}_S\right),
\end{align*}
where $\mbf{Q}_S=\mbf{M}^\prime\mbf{Q}\mbf{M}$. This implies $p+q+1$ model parameters.

This sparse model is more closely analogous to the centered automodel than is the RHZ model. The uncentered automodel accounts for the underlying graph, since the uncentered autocovariate is a sum over neighbors, but does not fit structure residual to $\mbf{X}$. The RHZ model accounts for $\mbf{X}$ but not for the underlying graph. Both the centered automodel and our sparse SGLMM account for $\mbf{X}$ and for the underlying graph. Table~\ref{modtab} compares and contrasts the five models treated here. 

\begin{table}
\caption{\label{modtab}Comparing and contrasting various spatial generalized linear models.}
\centering
\begin{tabular}{|ccccc|}\hline
Model & Systematic Distortion of Mean & Confounded & Accounts for $\mbf{X}$ & Accounts for $G$\\\hline\hline
Uncentered Automodel & $\eta\sum_{(i,j)\in E}Z_j$ & Yes & No & Yes\\
Traditional SGLMM & $\mbf{K}_i\bs{\gamma}+\mbf{L}_i\bs{\delta}$ & Yes & No & No\\
RHZ SGLMM & $\mbf{L}_i\bs{\delta}$ & No & Yes & No\\
Centered Automodel & $\eta\sum_{(i,j)\in E}(Z_j-\mu_j)$ & No & Yes & Yes\\
Sparse SGLMM & $\mbf{M}_i\bs{\delta}_S$ & No & Yes & Yes\\
\hline
\end{tabular}
\end{table}

We note that another popular measure of spatial dependence is Geary's $C$, which uses the graph Laplacian, $\mbf{Q}$, in place of $\mbf{A}$ \citep{Geary:1954p894}. Geary's $C$ is the spatial analog of the well-known Durbin-Watson statistic for measuring autocorrelation in the residuals from a time series regression \citep{Durbin:1950p893}. More specifically, the Durbin-Watson statistic is similar to Geary's $C$ for a path graph (where adjacency is in time rather than space). Moran's $I$ corresponds to a product-moment formulation, Geary's $C$ to a squared-difference formulation. The eigensystem of $\mbf{P}^\perp\mbf{Q}\mbf{P}^\perp$ is a viable alternative to that of $\mbf{P}^\perp\mbf{A}\mbf{P}^\perp$ for the current application, and perhaps other matrix representations of $G$ would also provide suitable eigensystems. We chose $\mbf{A}$ because the spectrum of a Moran operator is particularly easy to interpret.


\section{Dimension Reduction for Spatial Models}
\label{reduction}

Fitting a Gaussian process model like that mentioned above in Section~\ref{traditional} requires the repeated evaluation of expressions involving the inverse of the covariance matrix, $\mbf{H}(\bs{\phi})$, where $\bs{\phi}$ are the parameters of the spatial covariance function. The customary approach to this problem is to avoid inversion in favor of Cholesky decomposition of $\mbf{H}$ followed by a linear solve. Since $\mbf{H}$ is typically dense, its Cholesky decomposition is $\Theta(n^3)$, and so the time complexity of the overall fitting algorithm is $\Theta(n^3)$. This considerable computational expense makes the analyses of large point-level datasets time consuming or infeasible. Consequently, efforts to reduce the computational burden have resulted in an extensive literature detailing many approaches, e.g., process convolution \citep{Higdon:2002p909}, fixed rank kriging \citep{Cressie:2008p911}, Gaussian predictive process models \citep{Banerjee:2008p903}, covariance tapering \citep{Furrer:2006p912}, and approximation by a Gaussian Markov random field \citep{Rue:2002p915}.

Fitting an areal mixed model can also require expensive matrix operations. It is well known that a univariate Metropolis-Hastings algorithm for sampling from the posterior distribution of $\bs{W}$ leads to a slow mixing Markov chain because the components of $\bs{W}$ exhibit strong \emph{a posteriori} dependence. This has led to a number of approaches that involve updating the random effects in a block(s). Constructing proposals for these updates is challenging, and the improved mixing comes at the cost of increased running time per iteration \citep[see, for instance,][]{KnorrHeld:2002p267,Haran:2003p921,Haran:2010p922}.

The random effects for the RHZ model and for our sparse model, on the other hand, are practically \emph{a posteriori} uncorrelated. This means we can use a spherical normal proposal for the random effects, which is very efficient computationally. Thus the computational crux of fitting the RHZ (sparse) model  is the evaluation of the quadratic form $\bs{\delta}^\prime\mbf{Q}_R\bs{\delta}$ ($\bs{\delta}_S^\prime\mbf{Q}_S\bs{\delta}_S$). This operation has time complexity $\Theta(n)$ for the RHZ model, which is sufficient to discourage or prevent the application of the model to large datasets. As we will show in the next section, evaluation of $\bs{\delta}_S^\prime\mbf{Q}_S\bs{\delta}_S$ can be $\Theta(1)$, which renders our sparse model applicable to even very large datasets.

\section{Simulation Study}
\label{simulation}

\begin{figure}
\centering
      \includegraphics[scale=.7]{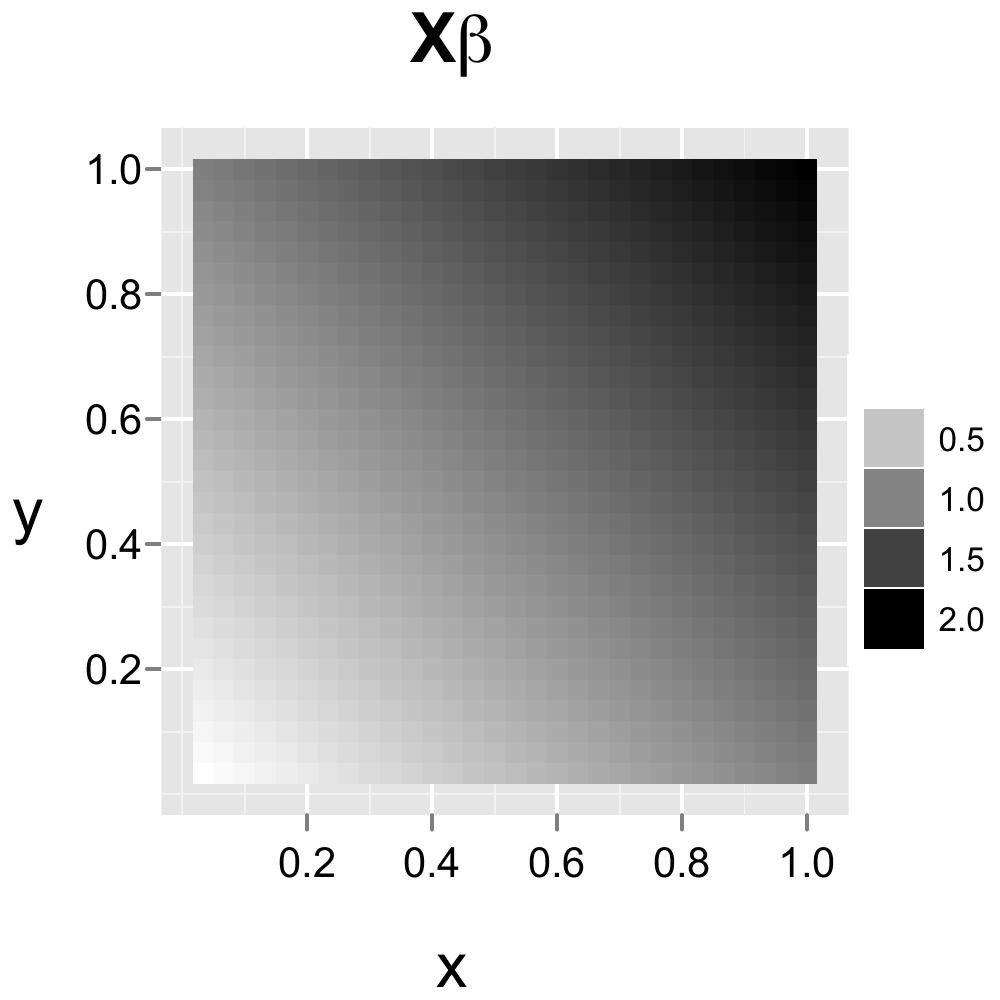}
\caption{\label{large}The large-scale structure, $\mbf{X}\bs{\beta}$, for our simulation study.}
\end{figure}

We applied the classical GLM and the three spatial models to binary, count, and normal data simulated from the sparse model. (We also conducted an extensive study wherein we applied these models to binary, count, and normal data simulated from the traditional SGLMM (with a proper GMRF). That study yielded similar results, which we have omitted in the interest of brevity.) The underlying graph for the binary and count data was the $30\times 30$ lattice, and we restricted the coordinates of the vertices to the unit square. We chose for our design matrix $\mbf{X}=[\bs{x}\;\bs{y}]$, where $\bs{x}=(x_1,\ldots,x_{900})^\prime$ and $\bs{y}=(y_1,\ldots,y_{900})^\prime$ are the $x$ and $y$ coordinates of the vertices, and we let $\bs{\beta}=(1,1)^\prime$. A level plot of this large-scale structure, $\mbf{X}\bs{\beta}=(x_1+y_1,\ldots,x_{900}+y_{900})^\prime$, is shown in Figure~\ref{large}.

\begin{figure}
\centering
      \includegraphics[scale=.4]{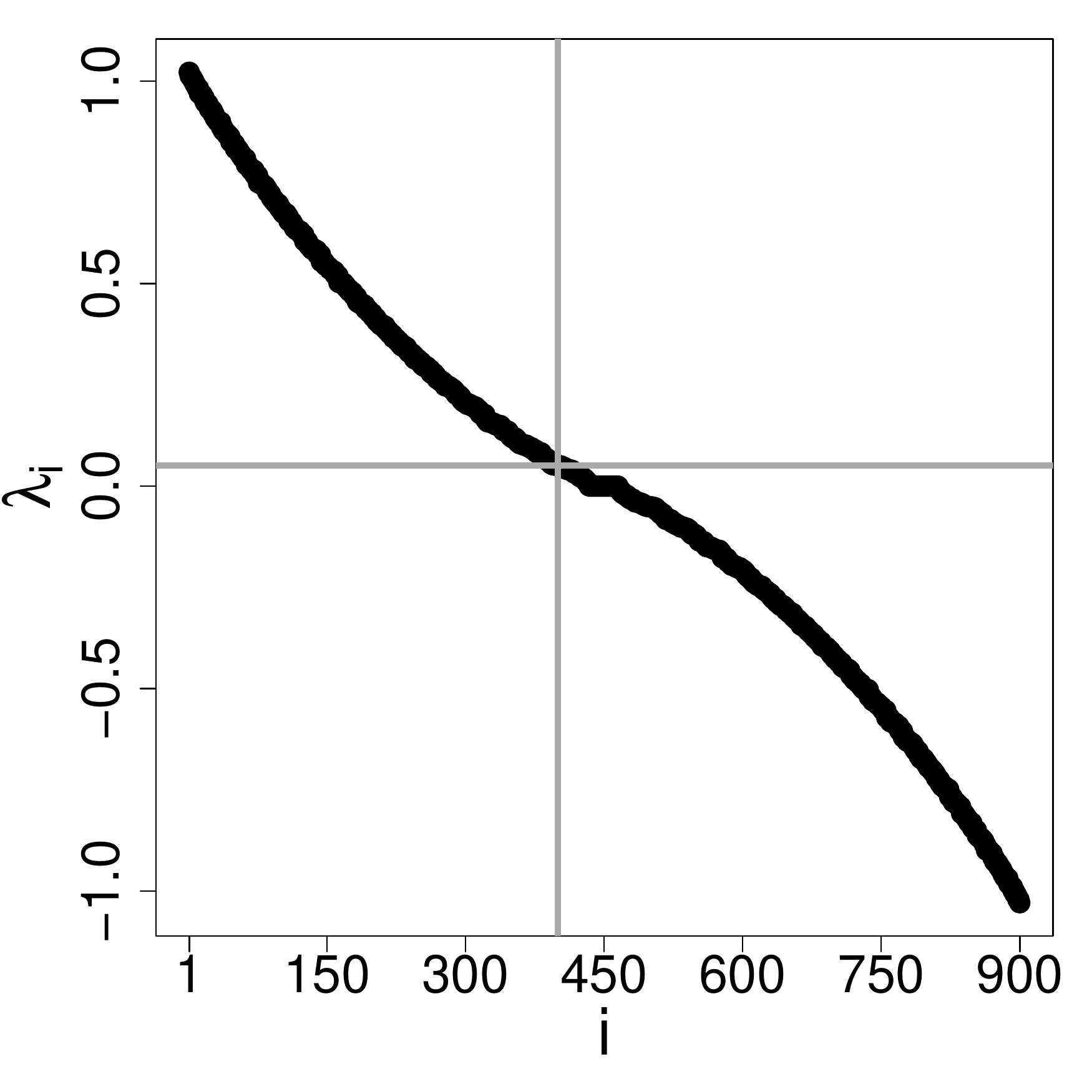}
\caption{\label{eigval}The standardized eigenvalues for the $30\times 30$ lattice.}
\end{figure}

The plot in Figure~\ref{eigval} shows the standardized eigenvalues for the $30\times 30$ lattice. Since over half of the values are non-positive, we chose to use only the first 400 eigenvectors to simulate data for our study, i.e., $\dim(\bs{\delta}_S)=400$ and $\mbf{M}$ is $900\times 400$. The horizontal gray line marks the 400th eigenvalue, which equals 0.05.

Since the traditional model requires a Bayesian analysis, we applied all three spatial models in a Bayesian setting. It is customary to put a flat prior or diffuse spherical normal prior on $\bs{\beta}$. We assumed that $\bs{\beta}\sim\nrm(\bs{0},100\,\mbf{I})$. The choice of a prior for $\tau$ has been the subject of debate. \citet{Kelsall1999Discussion-of-B} recommend a gamma prior with shape parameter 0.5 and scale parameter 2000. This prior is appealing because it corresponds to the prior belief that the fixed effects are sufficient to explain the data (since a large value for $\tau$ implies small variances for the random effects) and because it does not produce artifactual spatial structure in the posterior.

In all cases we simulated a sample path of length 2,000,000. We fit the SGLMMs using Metropolis-Hastings random walk updates and/or Gibbs updates. For the normal data we used Gibbs updates for all parameters, and we used a Gibbs update for $\tau$ for all three types of data. For the binary and count data we updated $\bs{\beta}$ using a random walk with proposal $\bs{\beta}^{(j+1)}\sim\nrm(\bs{\beta}^{(j)},\hat{\mbf{V}})$, where $\hat{\mbf{V}}$ is the estimated asymptotic covariance matrix from a classical GLM fit. We updated $\bs{W}$ using a univariate random walk with normal proposals. And we updated each of $\bs{\delta}$ and $\bs{\delta}_S$ using a multivariate random walk with a spherical normal proposal.

\subsection{Binary Data}

We created a binary dataset by first setting $\tau=1$ and simulating random effects according to $\bs{\delta}_S\sim\nrm(\bs{0},\mbf{Q}_S^{-1})$. Then we simulated independent observations according to $\bs{Z}\,|\,\bs{\delta}_S\sim\ber(\bs{p})$, where $\ber$ denotes a Bernoulli random variable and $\bs{p}=\exp(\bs{x}+\bs{y}+\mbf{M}\bs{\delta}_S)/(1 + \exp(\bs{x}+\bs{y}+\mbf{M}\bs{\delta}_S))$ is the vector of true probabilities.

\begin{figure}
\centering
\begin{tabular}{cc}
 \includegraphics[scale=.3]{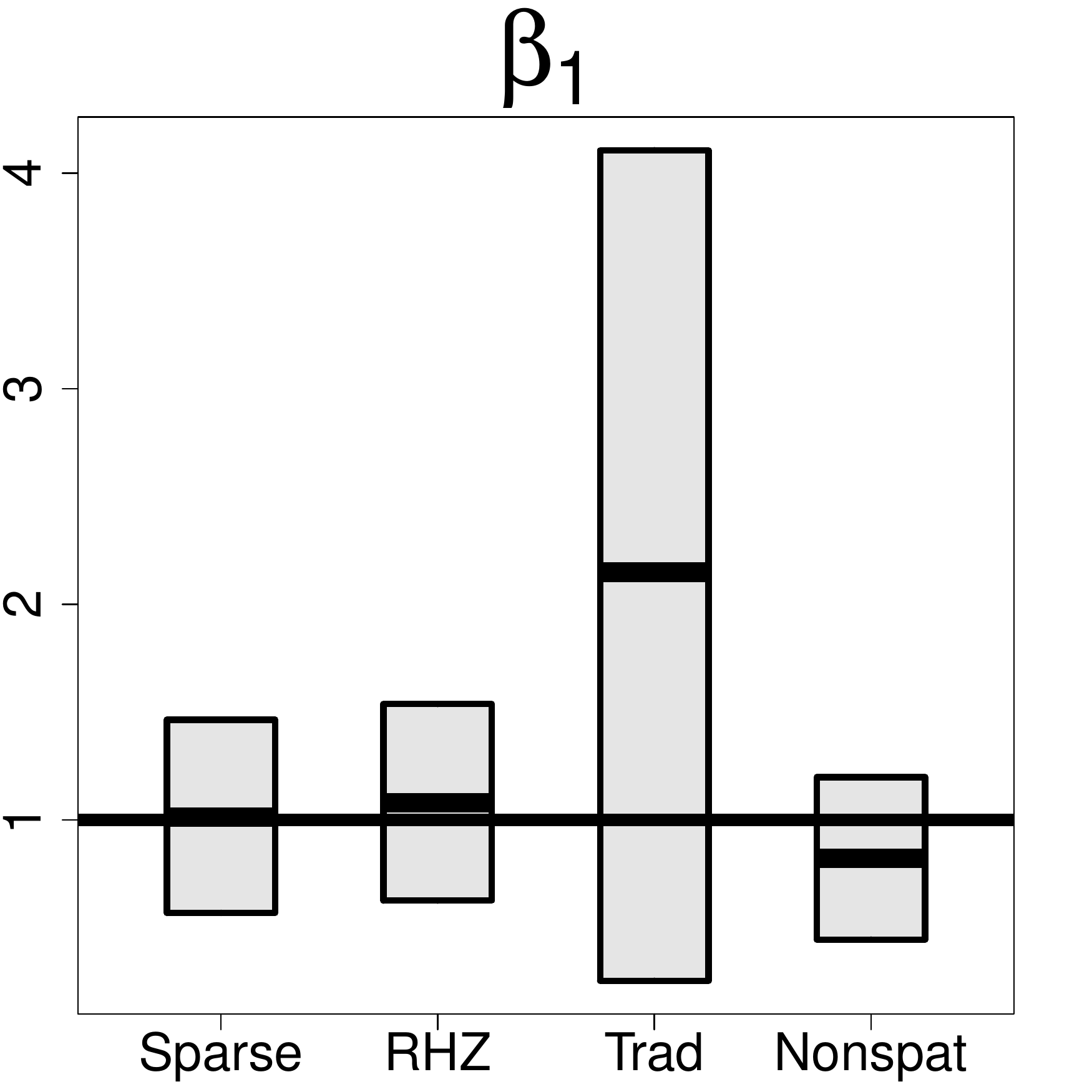} & \includegraphics[scale=.3]{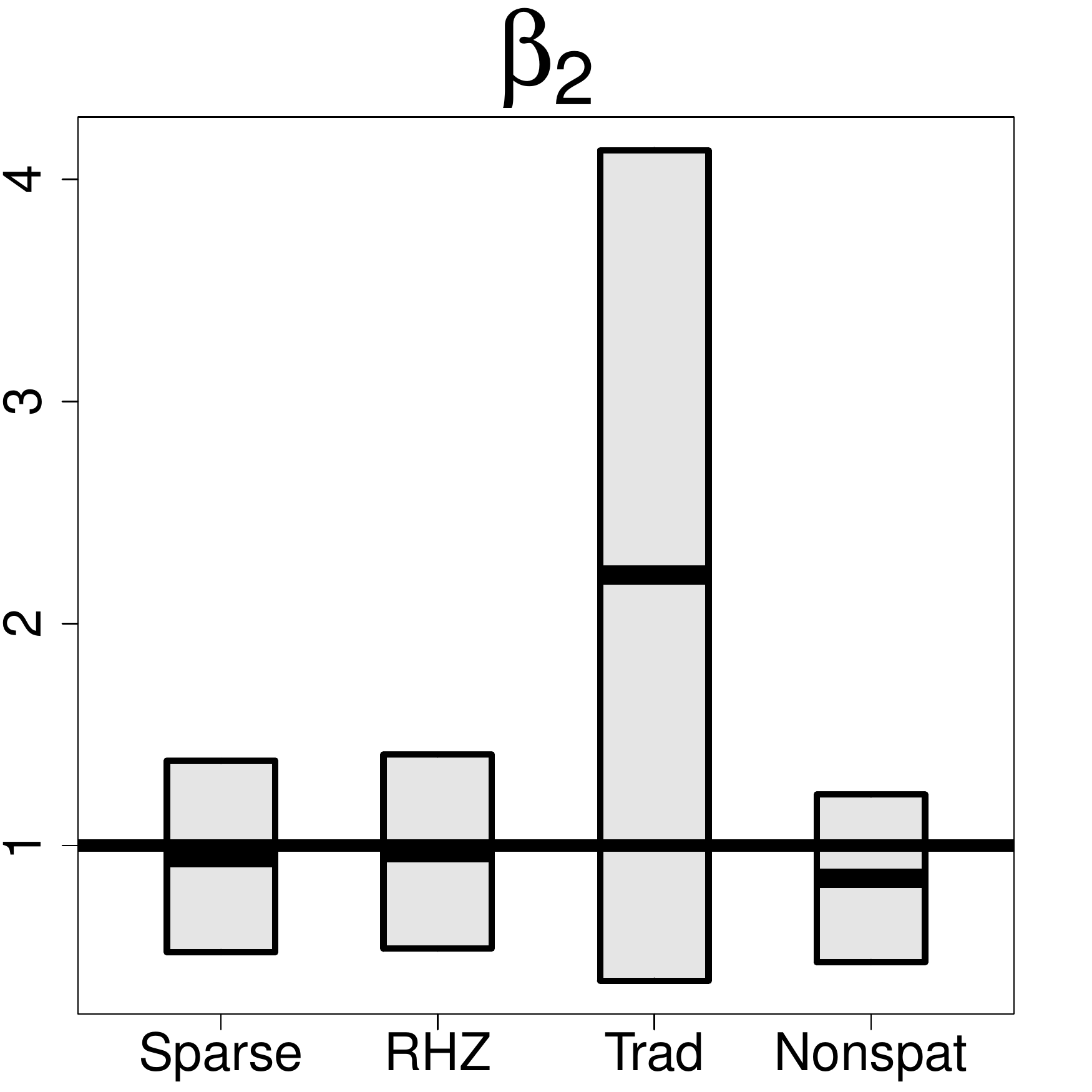}
\end{tabular}
   \caption{\label{boxbin}Box plots illustrating inference for $\bs{\beta}$ for the simulated binary data.}
\end{figure}

We fitted the simulated dataset using a nonspatial model, i.e., the standard logistic model; the centered autologistic model; the traditional SGLMM; the RHZ model; and our sparse model with varying degrees of sparsity---400 eigenvectors (the true model), 200 eigenvectors, 100, 50, and 25. The study results are shown in Table~\ref{simtabbinary}, and Figure~\ref{boxbin} illustrates the inference for $\bs{\beta}$ with box plots.

We see that the RHZ model and the true model produced approximately the same inference for these data. This is not surprising since the RHZ model is also capable of fitting residual structure and has random effects that essentially ``contain" the random effects for our sparse model, i.e., $\mbf{L}\bs{\delta}$ can accommodate any structure exhibited by $\mbf{M}\bs{\delta}_S$. The two models produced point estimates that are close to the true values, and confidence intervals that cover the true values but do not cover 0. The traditional SGLMM, on the other hand, gave a rather biased point estimate of $\bs{\beta}$ along with confidence intervals that are over four times as wide as those provided by the other SGLMMs.

The standard nonspatial logistic model gave a biased estimate of $\bs{\beta}$, which we expect of a model that does not account for dependence. But the nonspatial model's confidence intervals did at least cover the true value of $\bs{\beta}$ but not 0. The autologistic estimate was also biased, albeit less so, and it too gave valid confidence intervals, but those intervals are slightly wider than those for the other models (except the traditional SGLMM).

\begin{table}
\caption{\label{simtabbinary}The results of our simulation study for an SGLMM for binary data, i.e., for a Bernoulli first stage. Running times are given in minutes.}
\centering
\setlength{\extrarowheight}{1.5ex}
\begin{tabular}{|cccccccccc|}\hline
Model & Dim & $\hat{\beta}_1$ & CI($\beta_1$) & $\hat{\beta}_2$ & CI($\beta_2$) & $\hat{\tau}$ & CI($\tau$) & $\Vert\bs{p}-\hat{\bs{p}}\Vert$ & Time\\\hline\hline
Nonspatial & -- & 0.822 & (0.445, 1.199) & 0.852 & (0.475, 1.230) & -- & -- & 4.576 & --\\
Autologistic & -- & 0.930 & (0.446, 1.397) & 0.915 & (0.448, 1.367) & -- & -- & 3.876 & 20\\
Traditional & 900 & 2.149 & (0.254, 4.107) & 2.218 & (0.391, 4.132) & 2.803 & (0.287, 6.729) & 3.255 & 2,455\\
RHZ & 898 & 1.079 & (0.626, 1.538) & 0.969 & (0.536, 1.410) & 0.949 & (0.481, 1.482) & 3.263 & 1,640\\
Sparse & 400 & 1.013 & (0.569, 1.465) & 0.946 & (0.520, 1.382) & 1.425 & (0.355, 2.477) & 3.174 & 285\\
Sparse & 200 & 1.014 & (0.582, 1.452) & 0.937 & (0.519, 1.359) & 1.052 & (0.365, 1.872) & 3.175 & 167\\
Sparse & 100 & 0.997 & (0.566, 1.426) & 0.923 & (0.510, 1.342) & 0.935 & (0.339, 1.724) & 3.201 & 80\\
Sparse & 50 & 0.982 & (0.562, 1.403) & 0.900 & (0.495, 1.304) & 0.945 & (0.317, 1.751) & 3.295 & 49\\
Sparse & 25 & 0.964 & (0.552, 1.377) & 0.897 & (0.501, 1.301) & 0.867 & (0.225, 1.699) & 3.543 & 37\\
\hline
\end{tabular}
\end{table}

\begin{figure}
\centering
\begin{tabular}{cc}
 \includegraphics[scale=.3]{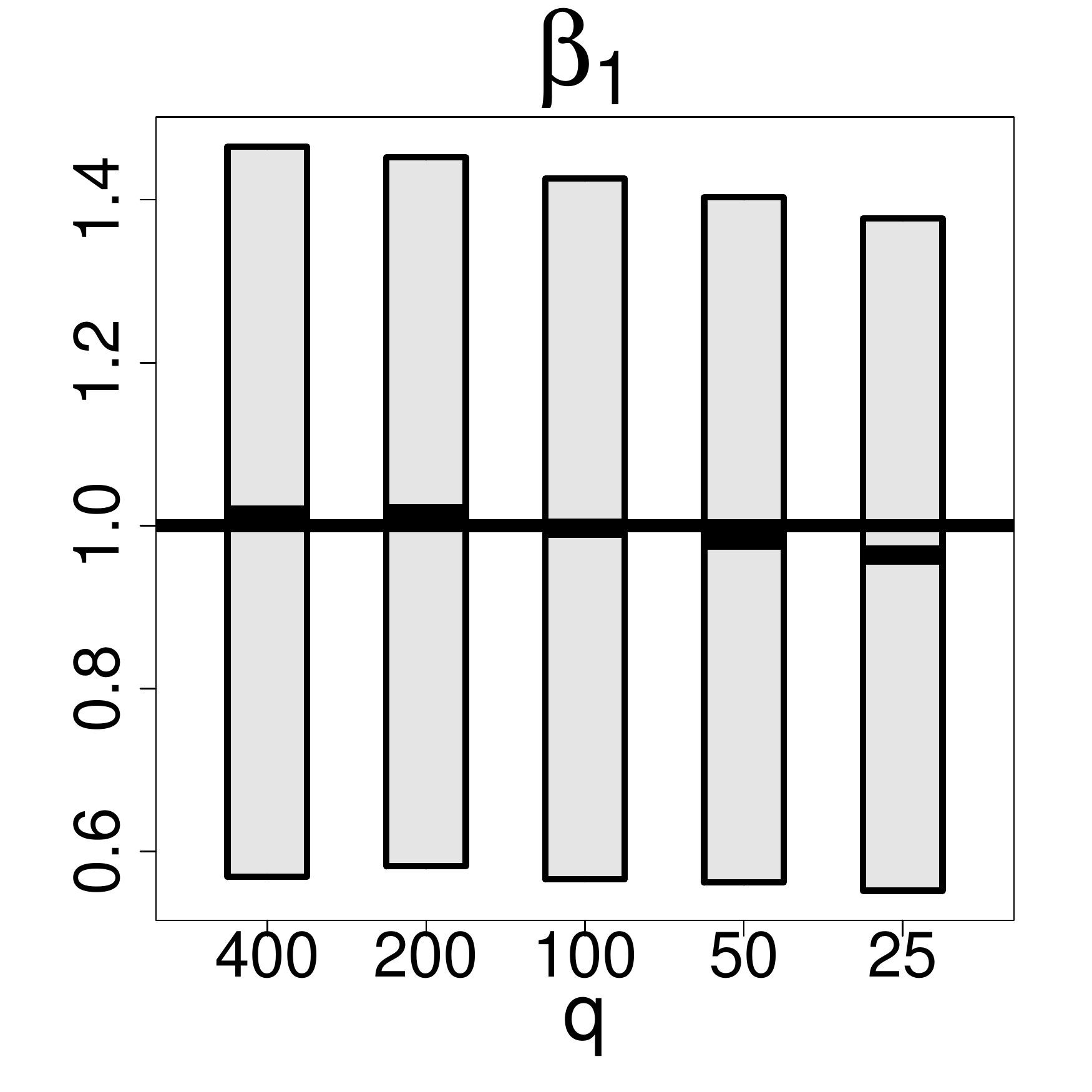} & \includegraphics[scale=.3]{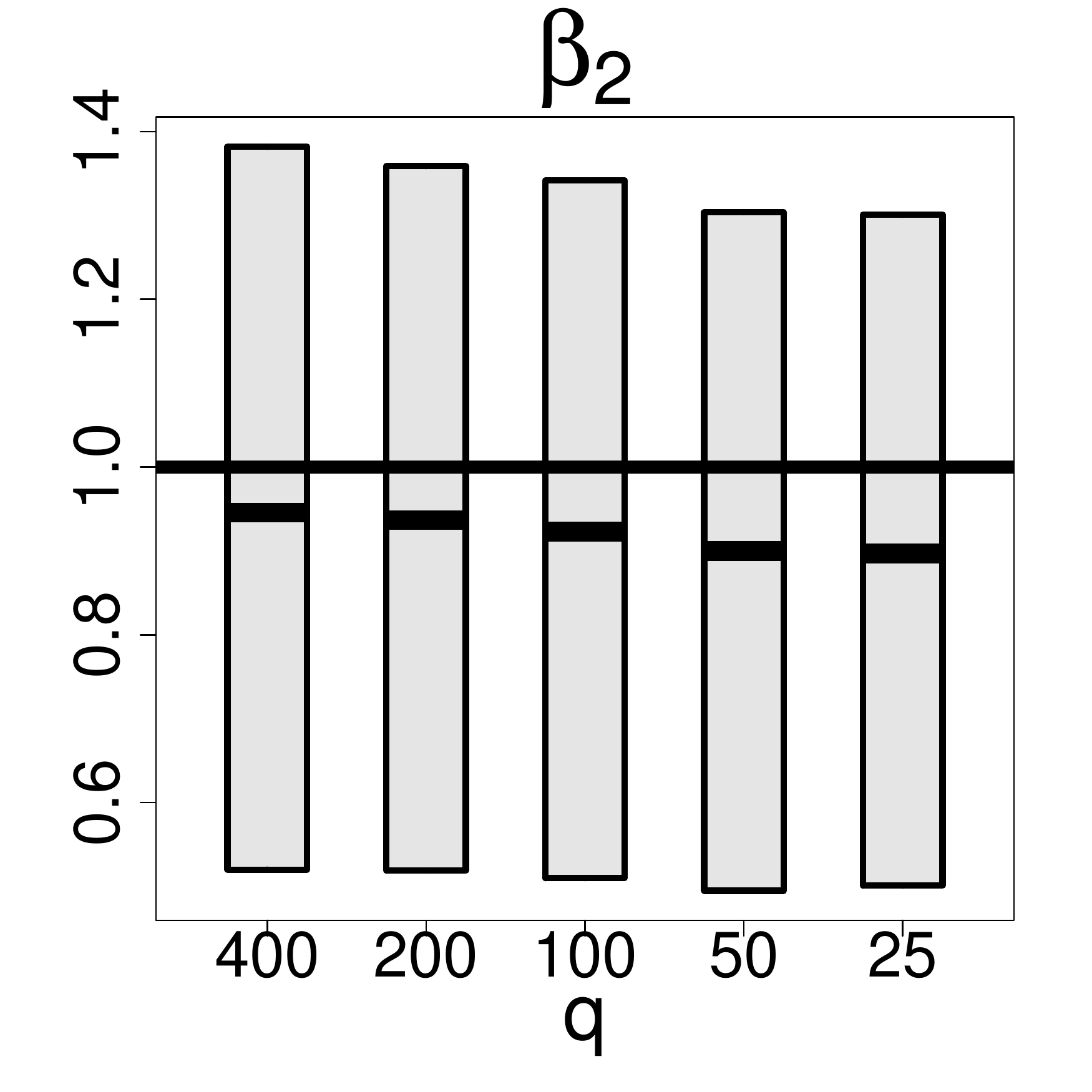}
\end{tabular}
   \caption{\label{boxsparse}Box plots illustrating the effect of increasing sparsity on regression inference for the binary data.}
\end{figure}

\begin{figure}
\centering
\begin{tabular}{ccc}
 \includegraphics[scale=.5]{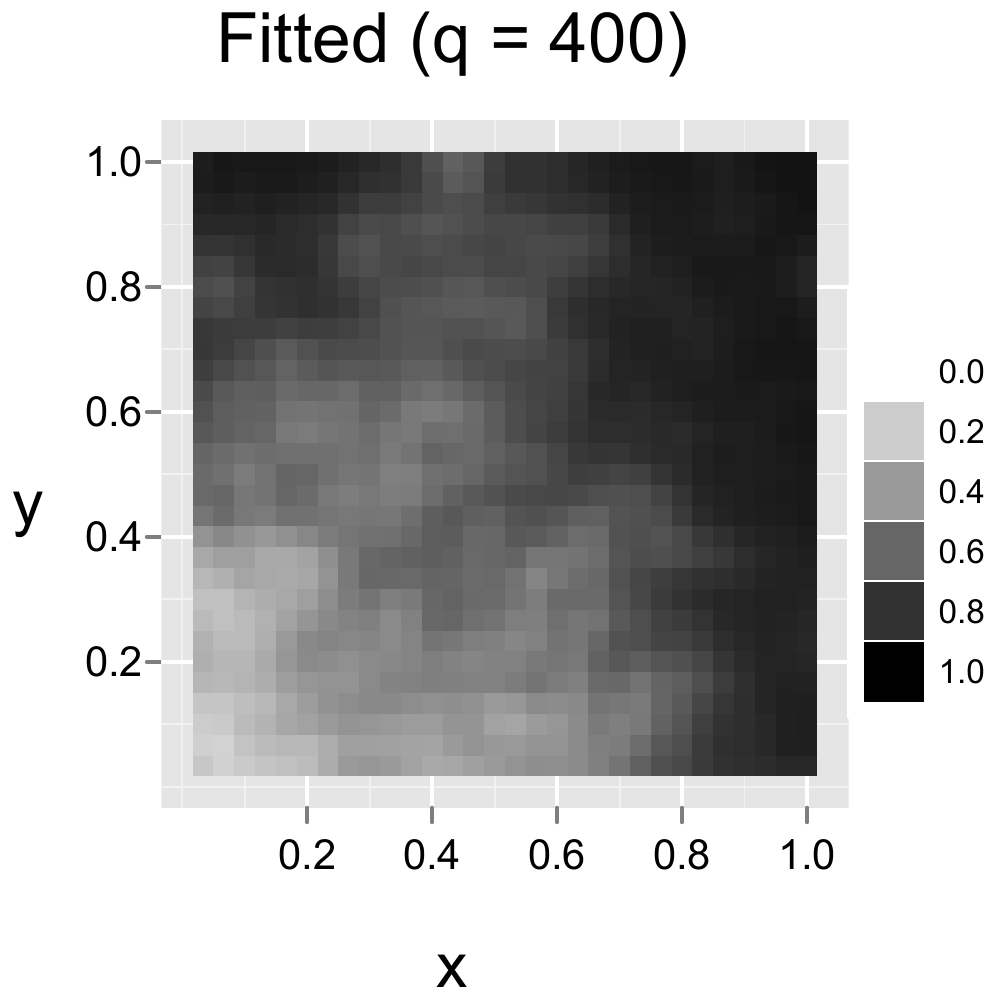} & \includegraphics[scale=.5]{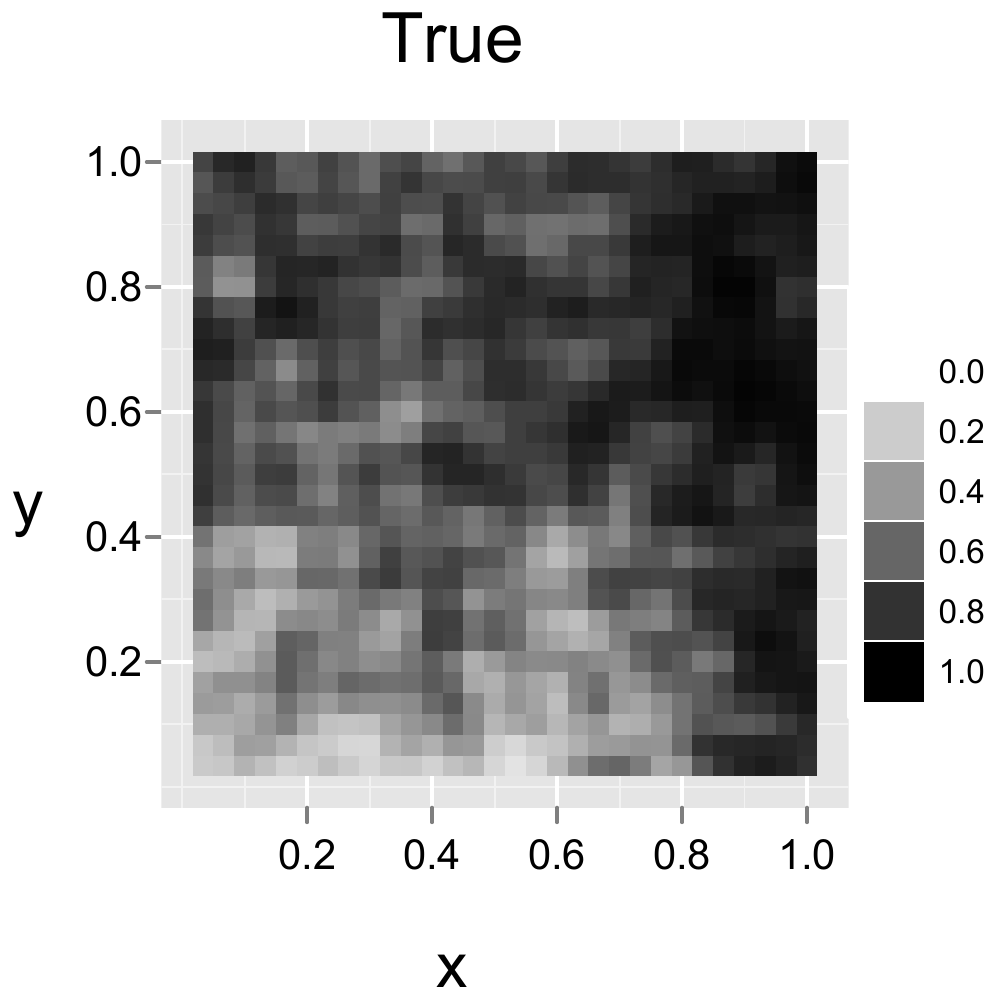} & \includegraphics[scale=.5]{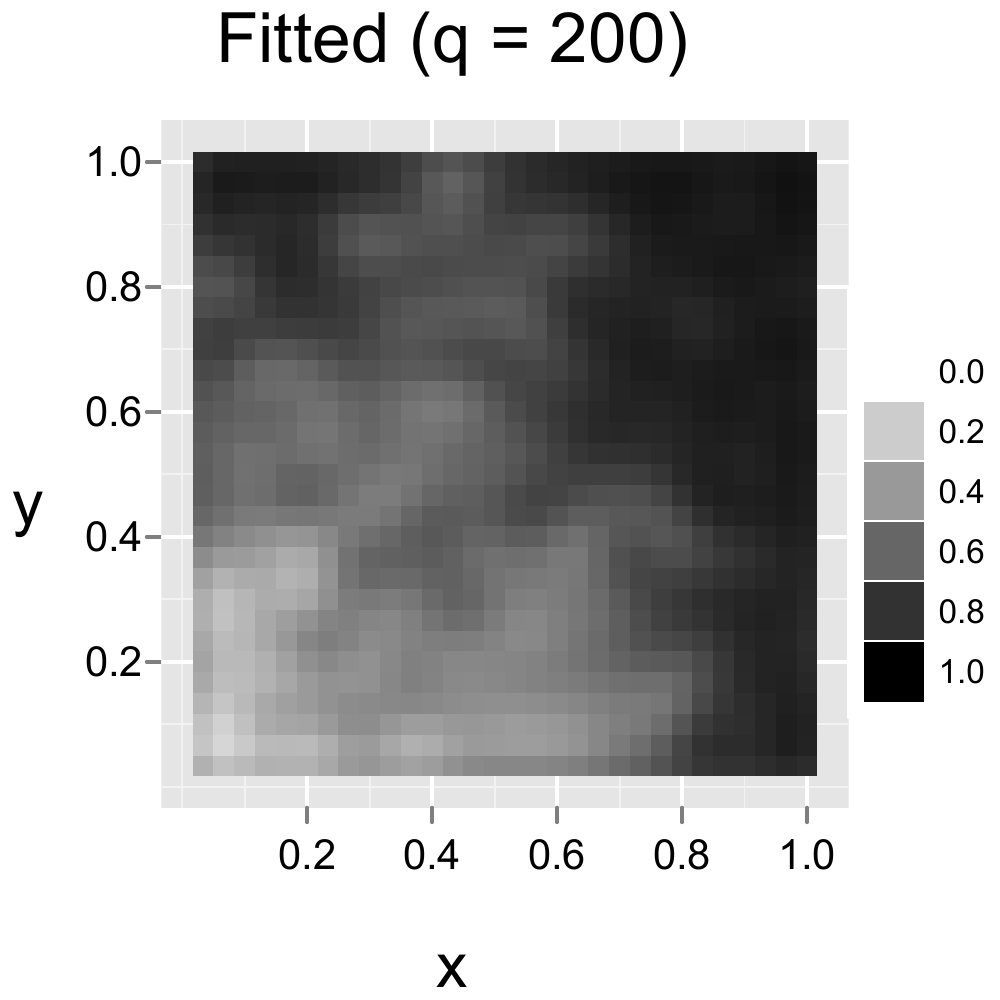}\\
 \includegraphics[scale=.5]{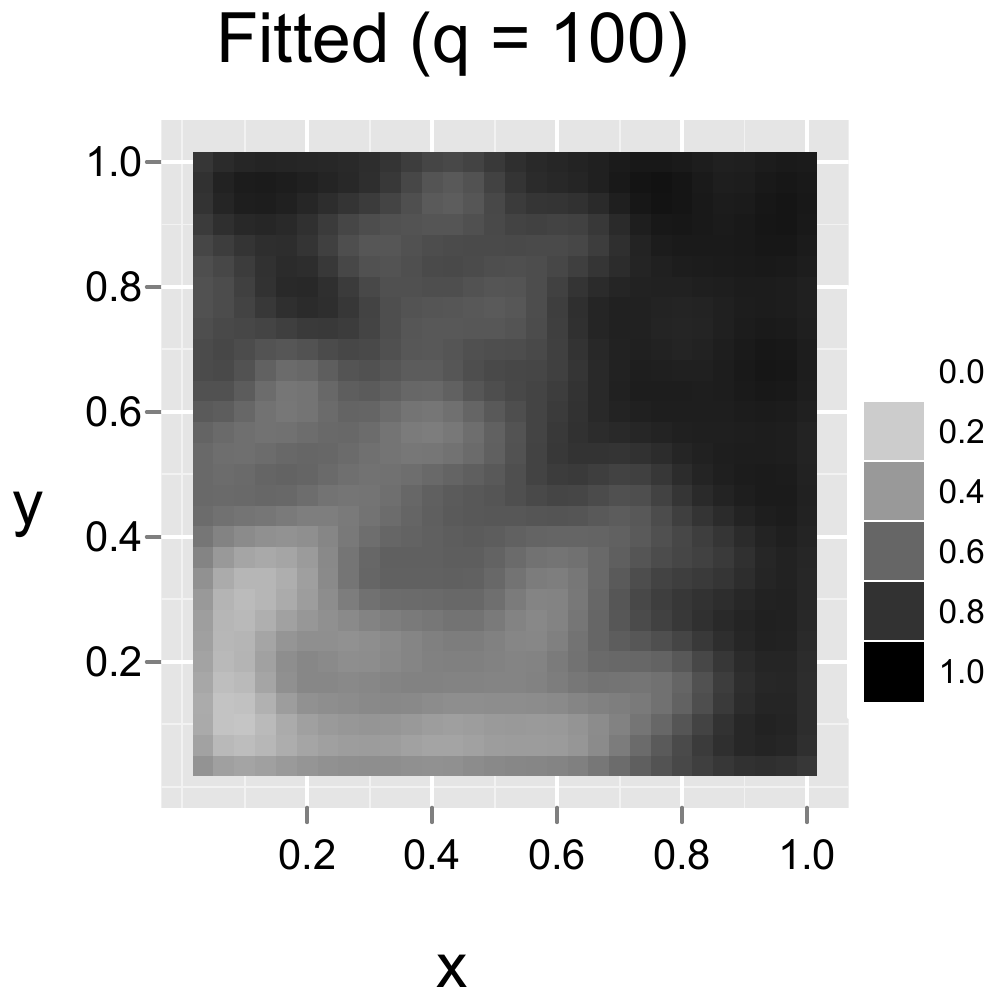} & \includegraphics[scale=.5]{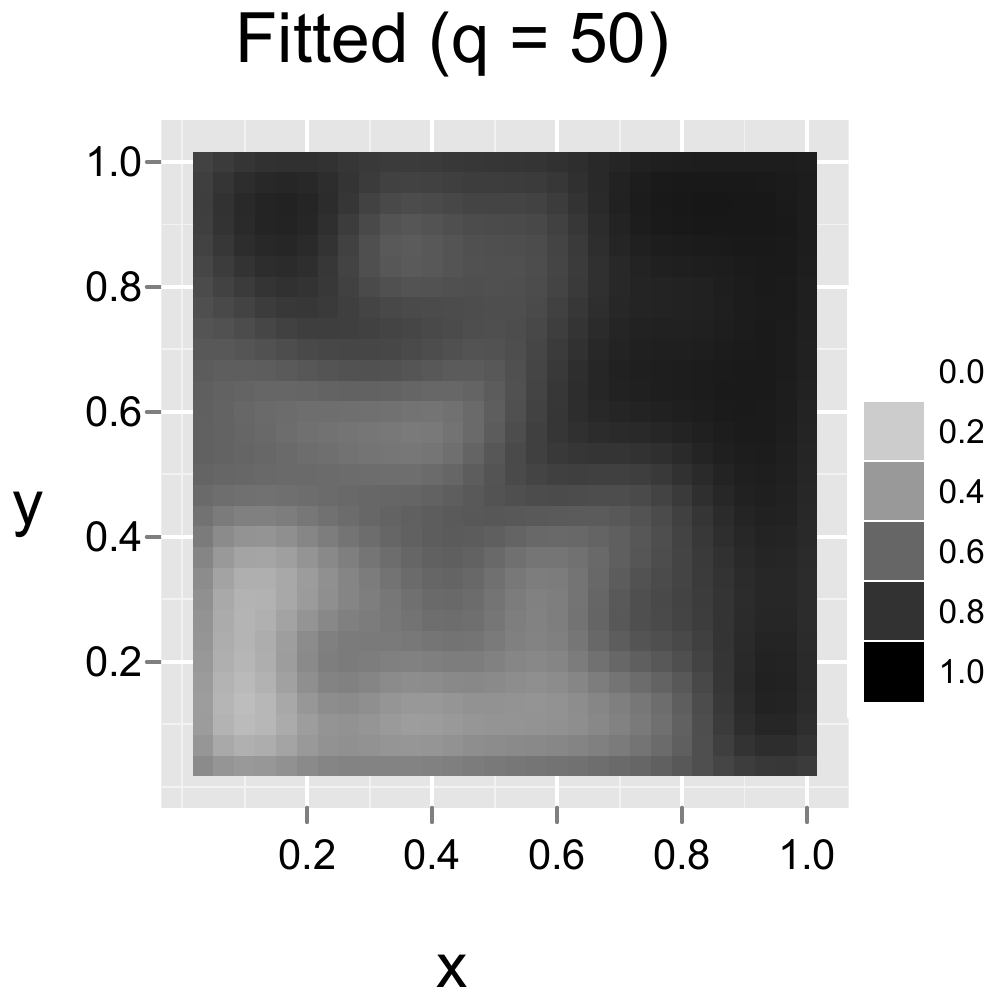} & \includegraphics[scale=.5]{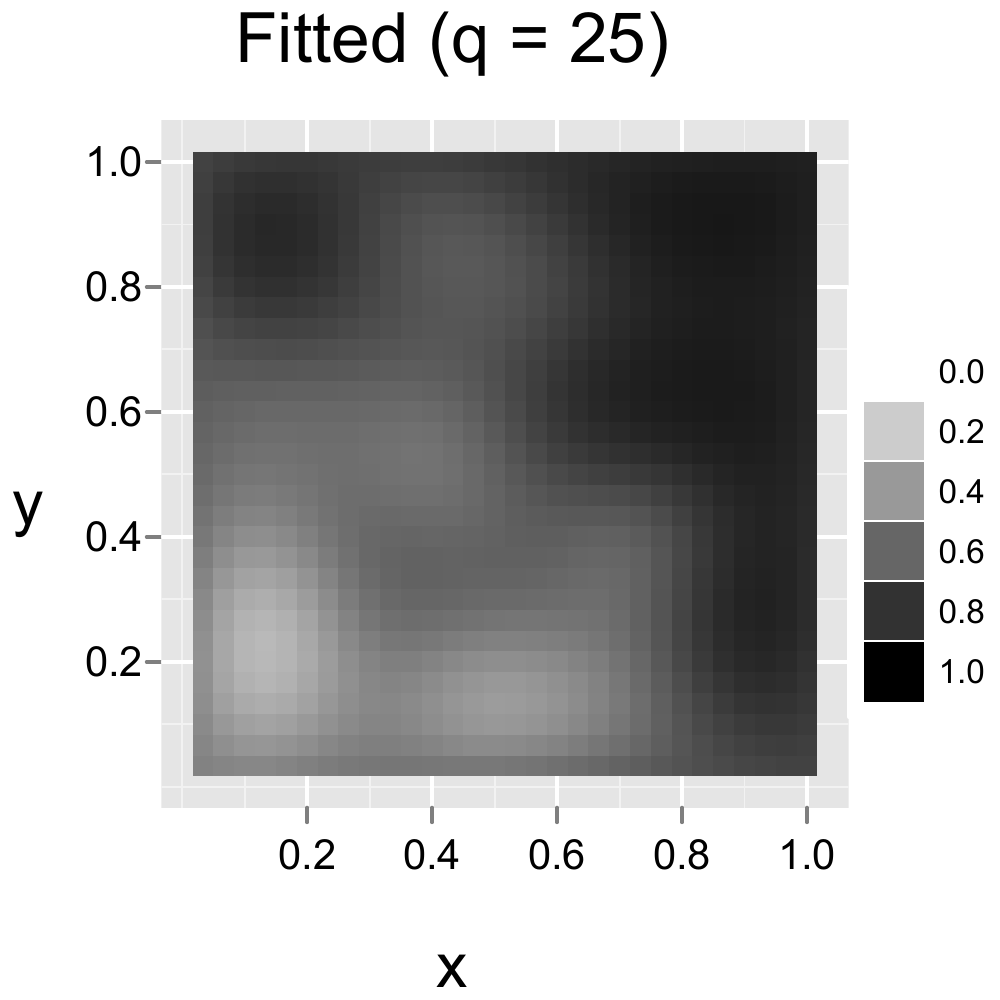}
\end{tabular}
   \caption{\label{p}The true and fitted probabilities for the simulated binary data, for various values of $q=\dim(\bs{\delta}_S)$.}
\end{figure}

The box plots in Figure~\ref{boxsparse} and the level plots in Figure~\ref{p} show the effect of increasing sparsity on regression inference and on fit, respectively. As we apply our sparse SGLMM with fewer and fewer eigenvectors, the posterior distribution of $\bs{\beta}$ exhibits increasing bias and decreasing variance, i.e., we observe the expected bias-variance tradeoff. The inference and fit provided by our model did not suffer appreciably until the number of eigenvectors had been reduced to 25, which represents a 97\% reduction in the number of parameters relative to the traditional and RHZ models.

\begin{figure}
\centering
 \includegraphics[scale=.4]{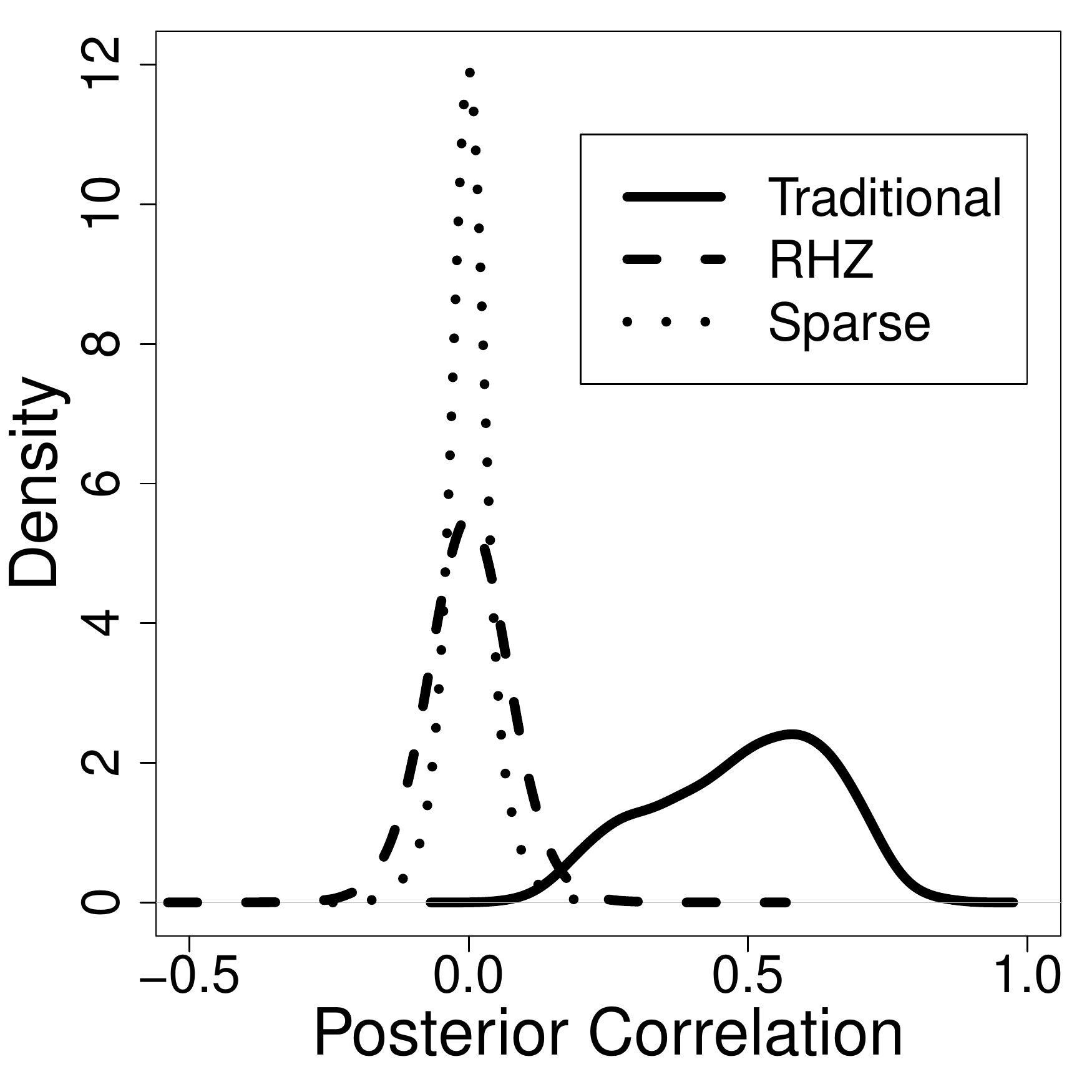}
   \caption{\label{poscor}Posterior correlations among the random effects for the three SGLMMs.}
\end{figure}

Figure~\ref{poscor} shows the distributions of the estimated posterior correlations among the random effects for the three SGLMMs. We see that the random effects for the RHZ model and for our sparse model are at most weakly correlated, while the random effects for the traditional model are often strongly dependent. This is why spherical normal proposals are sufficient for $\bs{\delta}$ and $\bs{\delta}_S$.

\subsection{Count Data}

We created a count dataset by first simulating random effects according to $\bs{\delta}_S\sim\nrm(\bs{0},(3\mbf{Q}_S)^{-1})$ and then simulating independent observations according to $\bs{Z}\,|\,\bs{\delta}_S\sim\poi(\bs{\lambda})$, where $\poi$ denotes a Poisson random variable and $\bs{\lambda}=\exp(\bs{x}+\bs{y}+\mbf{M}\bs{\delta}_S)$ is the vector of true rates.

The study results for the count data are shown in Table~\ref{simtabcount}. The highest posterior density intervals for all three spatial models cover the true value of $\bs{\beta}$, while the standard Poisson regression model yielded erroneous inference for $\beta_2$ because the model's failure to account for small-scale structure resulted in a poor point estimate. The traditional SGLMM produced a good point estimate of $\bs{\beta}$ for this dataset but once again gave much wider intervals than the other models.

\begin{table}
\caption{\label{simtabcount}The results of our simulation study for an SGLMM for count data, i.e., for a Poisson first stage. Running times are given in minutes.}
\centering
\setlength{\extrarowheight}{1.5ex}
\begin{tabular}{|cccccccccc|}\hline
Model & Dim & $\hat{\beta}_1$ & CI($\beta_1$) & $\hat{\beta}_2$ & CI($\beta_2$) & $\hat{\tau}$ & CI($\tau$) & $\Vert\bs{\lambda}-\hat{\bs{\lambda}}\Vert$ & Time\\\hline\hline
Nonspatial & -- & 1.108 & (0.989, 1.226) & 1.179 & (1.061, 1.297) & -- & -- & 53.898 & --\\
Traditional & 900 & 1.007 & (0.073, 1.913) & 1.056 & (0.117, 1.986) & 5.141 & (3.469, 6.952) & 26.039 & 2,275\\
RHZ & 898 & 0.934 & (0.816, 1.051) & 1.092 & (0.975, 1.210) & 4.978 & (2.916, 7.324) & 26.172 & 1,864\\
Sparse & 400 & 0.942 & (0.824, 1.060) & 1.101 & (0.986, 1.215) & 4.499 & (2.873, 6.402) & 24.379 & 485\\
\hline
\end{tabular}
\end{table}

\begin{figure}
\centering
\begin{tabular}{cc}
 \includegraphics[scale=.3]{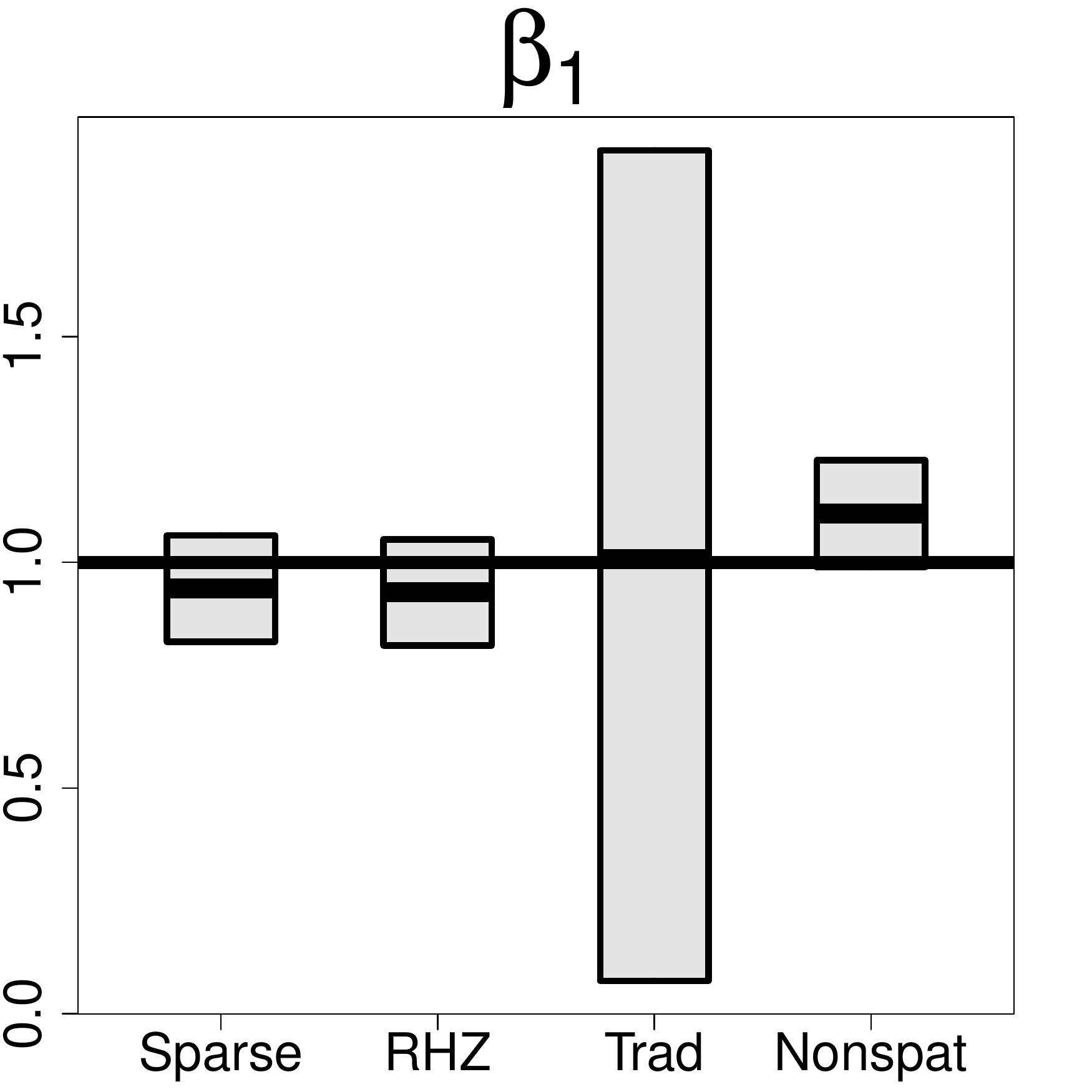} & \includegraphics[scale=.3]{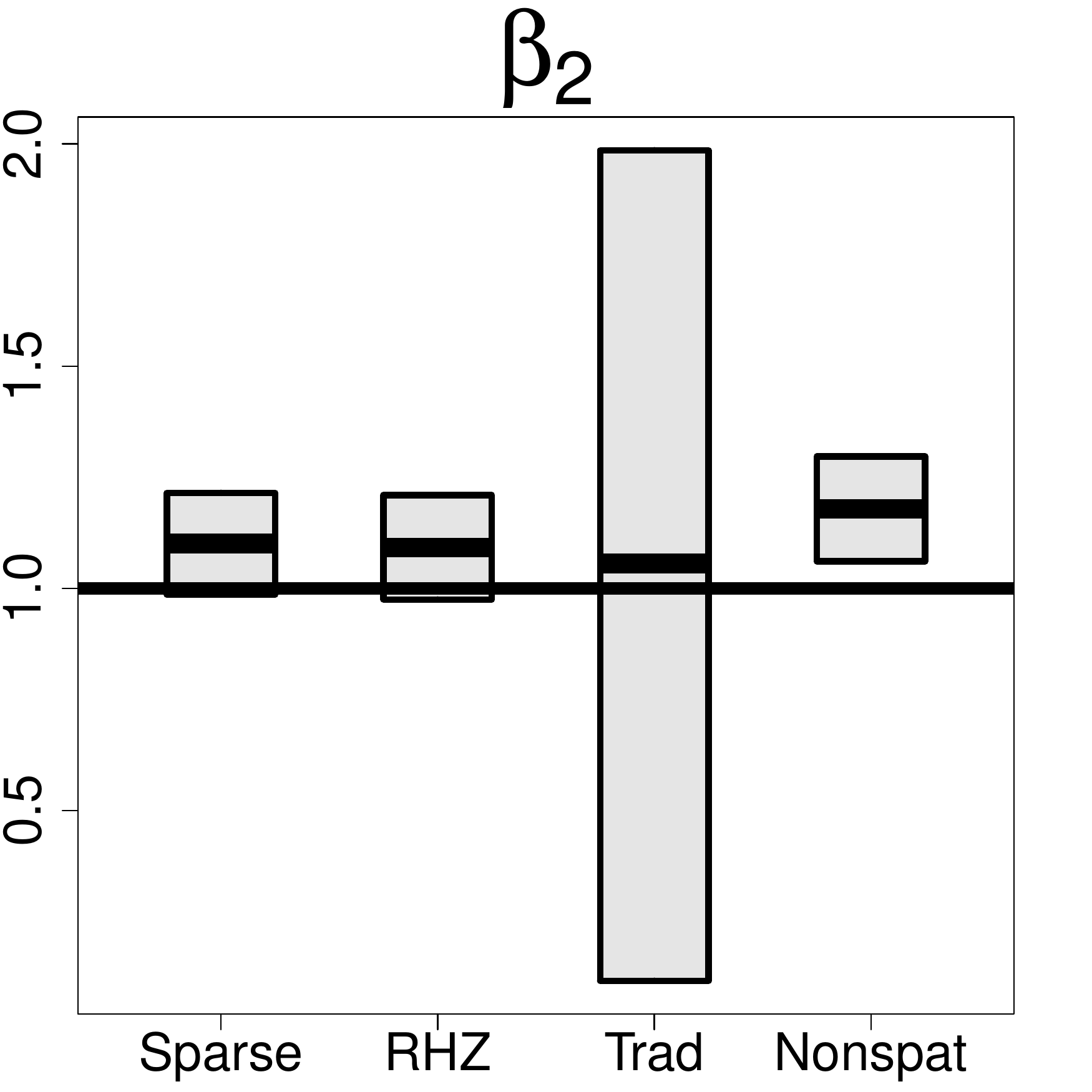}
\end{tabular}
   \caption{\label{boxcnt}Box plots illustrating inference for $\bs{\beta}$ for the simulated count data.}
\end{figure}

\subsection{Normal Data}

For the normal data we used the $20\times 20$ lattice with the coordinates of the vertices again restricted to the unit square, and we used the first 180 eigenvectors of the corresponding Moran operator. We simulated random effects according to $\bs{\delta}_S\sim\nrm(\bs{0},\mbf{Q}_S^{-1})$ and then simulated observations according to $\bs{Z}\,|\,\bs{\delta}_S\sim\nrm(\bs{\mu}=\bs{x}+\bs{y}+\mbf{M}\bs{\delta}_S, \sigma^2\mbf{I})$, where $\sigma^2=1$.

The study results for these data are shown in Table~\ref{simtabnorm}. The RHZ and sparse models gave narrower intervals than the ordinary linear model because the latter model overestimated $\sigma^2$. For this dataset the traditional SGLMM gave a poor estimate of $\bs{\beta}$ and inflated the variance so much as to cause a Type II error. We note that the running times for this study were much longer than they would have been for binary and count datasets of the same size because we did a Gibbs update of the random effects for the normal data, which requires an expensive Cholesky decomposition.

\begin{table}
\caption{\label{simtabnorm}The results of our simulation study for an SLMM, i.e., for a Gaussian first stage. Running times are given in minutes.}
\centering
\setlength{\extrarowheight}{1.5ex}
\begin{tiny}
\begin{tabular}{|cccccccccccc|}\hline
Model & Dim & $\hat{\beta}_1$ & CI($\beta_1$) & $\hat{\beta}_2$ & CI($\beta_2$) & $\hat{\tau}$ & CI($\tau$) & $\hat{\sigma}^2$ & CI($\sigma^2$) & $\Vert\bs{\mu}-\hat{\bs{\mu}}\Vert$ & Time\\\hline\hline
Nonspatial & -- & 1.143 & (0.824, 1.462) & 0.925 & (0.606, 1.244) & -- & -- & 1.558 & -- & 14.204 & --\\
Traditional & 400 & 1.583 & (-0.816, 3.984) & 0.288 & (-2.066, 2.656) & 0.825 & (0.417, 1.322) & 0.793 & (0.585, 1.312) & 9.798 & 1,248\\
RHZ & 398 & 1.143 & (0.912, 1.377) & 0.925 & (0.692, 1.158) & 0.847 & (0.427, 1.365) & 0.802 & (0.595, 1.141) & 9.725 & 1,660\\
Sparse & 180 & 1.143 & (0.884, 1.403) & 0.925 & (0.666, 1.185) & 0.961 & (0.525, 1.486) & 1.020 & (0.866, 1.229) & 8.692 & 385\\
\hline
\end{tabular}
\end{tiny}
\end{table}

\begin{figure}
\centering
\begin{tabular}{cc}
 \includegraphics[scale=.3]{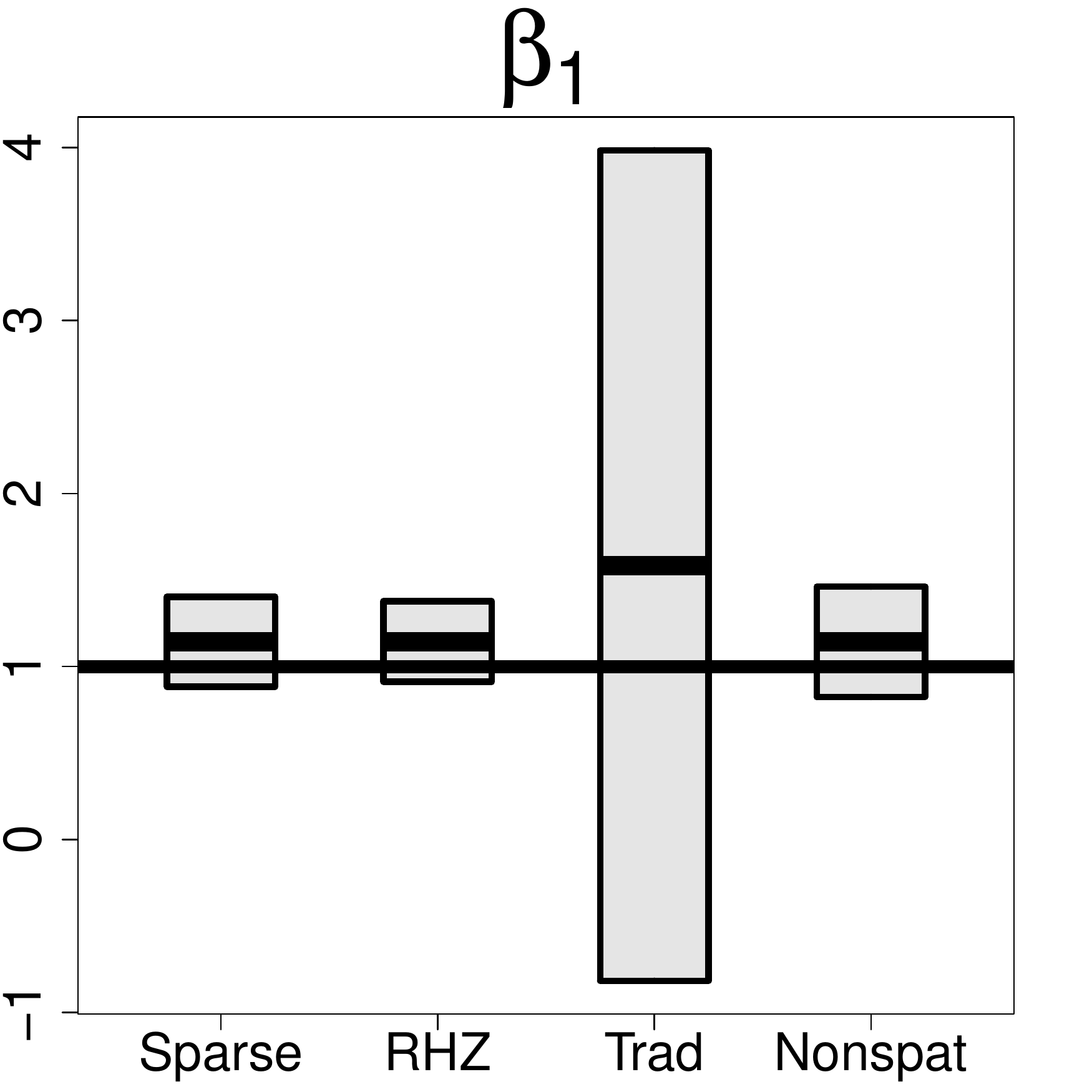} & \includegraphics[scale=.3]{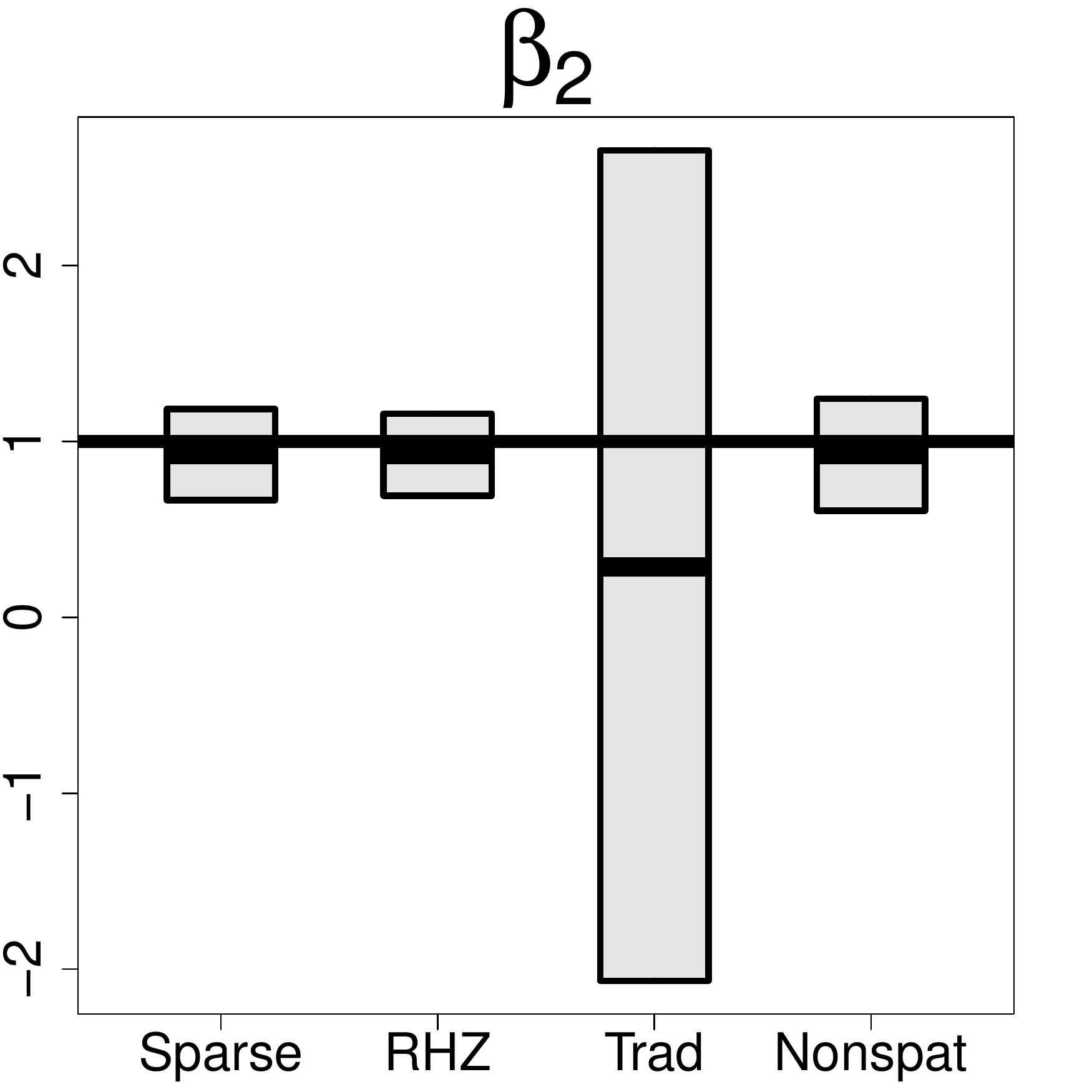}
\end{tabular}
   \caption{\label{boxnrm}Box plots illustrating inference for $\bs{\beta}$ for the simulated Gaussian data.}
\end{figure}

\subsection{A Heuristic for Dimension Reduction}

We conducted a more extensive simulation study with the aim of providing a heuristic for unsupervised dimension reduction, i.e., dimension reduction that does not employ the data to be analyzed. For this study we used the $50 \times 50$ lattice, which is large but not so large as to make fitting the true model infeasible. We used the same regression component as before, i.e., $\mbf{X}=[\bs{x}\,\bs{y}]$ and $\bs{\beta}=(1,1)^\prime$, and simulated binary datasets for four values of the smoothing parameter, $\tau$: 0.5, 1, 2, and 4. We used the first 1,100 eigenvectors of the Moran operator, which correspond to standardized eigenvalues greater than 0.06 and thus practically exhaust the patterns of positive dependence for this scenario. We then fit the true model and four sparser models to the simulated data. The number of random effects for the sparser models were 625 (the upper quartile of the eigenvalues), 265 (standardized eigenvalues greater than 0.7), 100, and 50. The last two of these were chosen with the hope of recommending a small, fixed dimension for the random effects.

\begin{figure}
\centering
\begin{tabular}{cccc}
$\tau=0.5$ & $\tau=1$ & $\tau=2$ & $\tau=4$ \\
 \includegraphics[scale=.2]{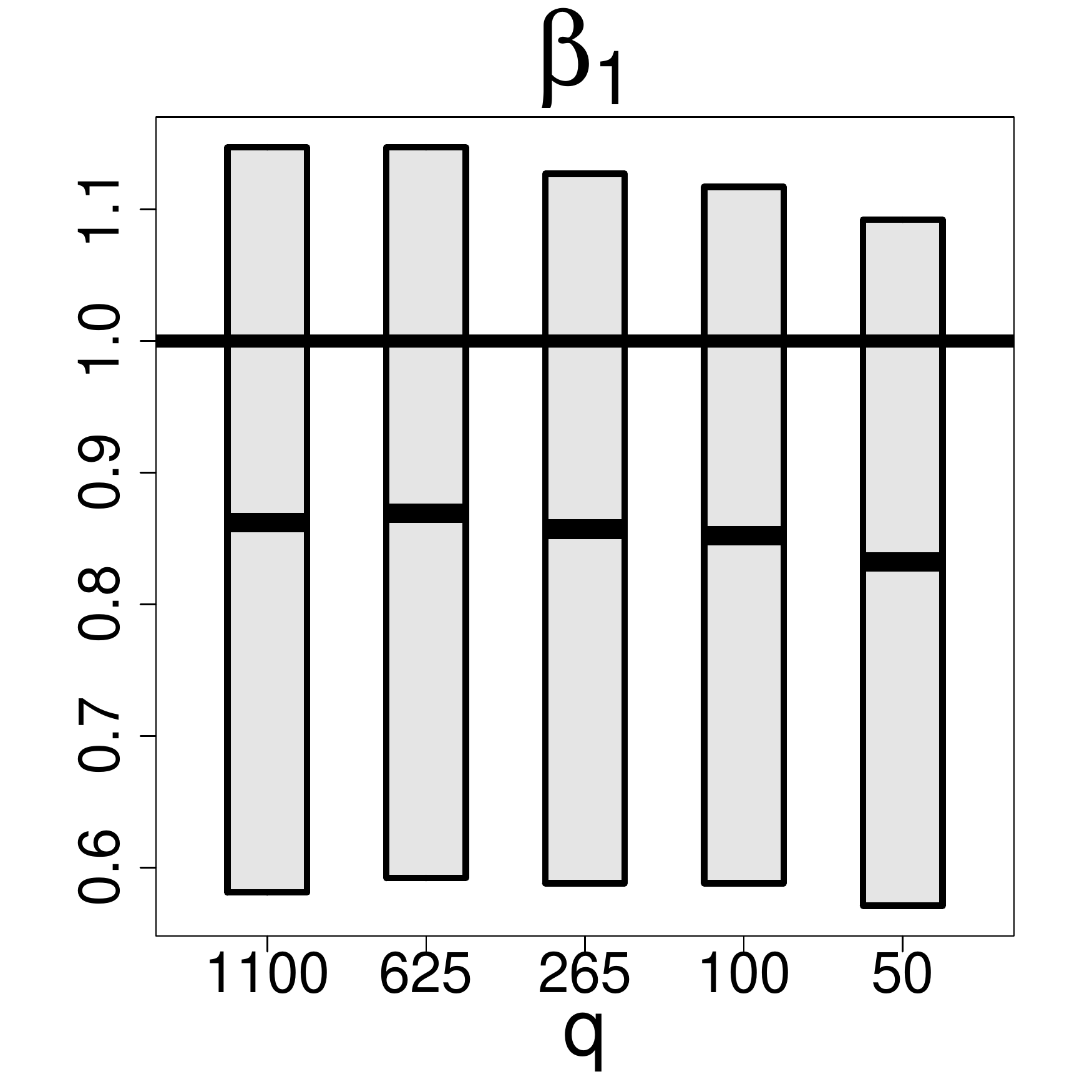} & \includegraphics[scale=.2]{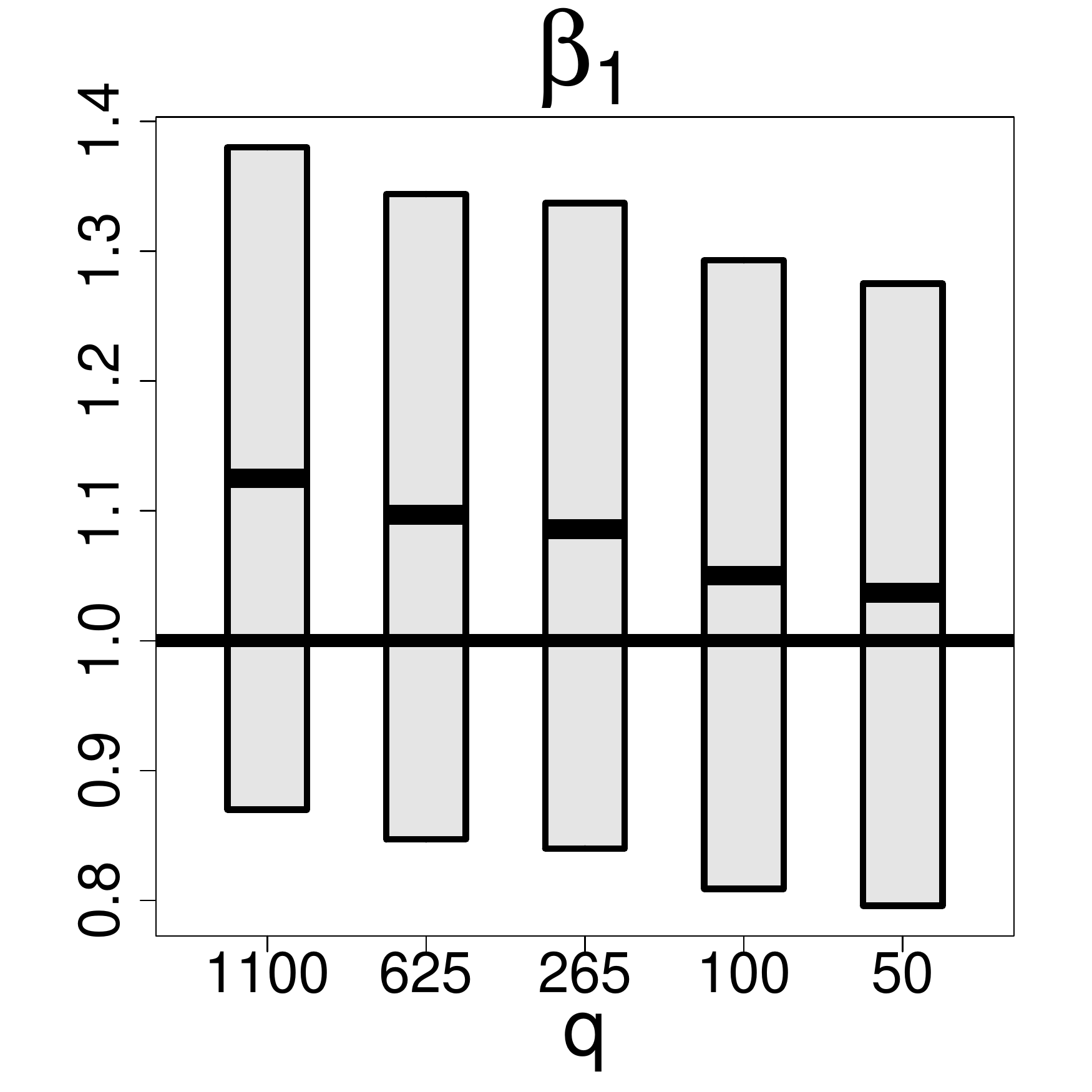} & \includegraphics[scale=.2]{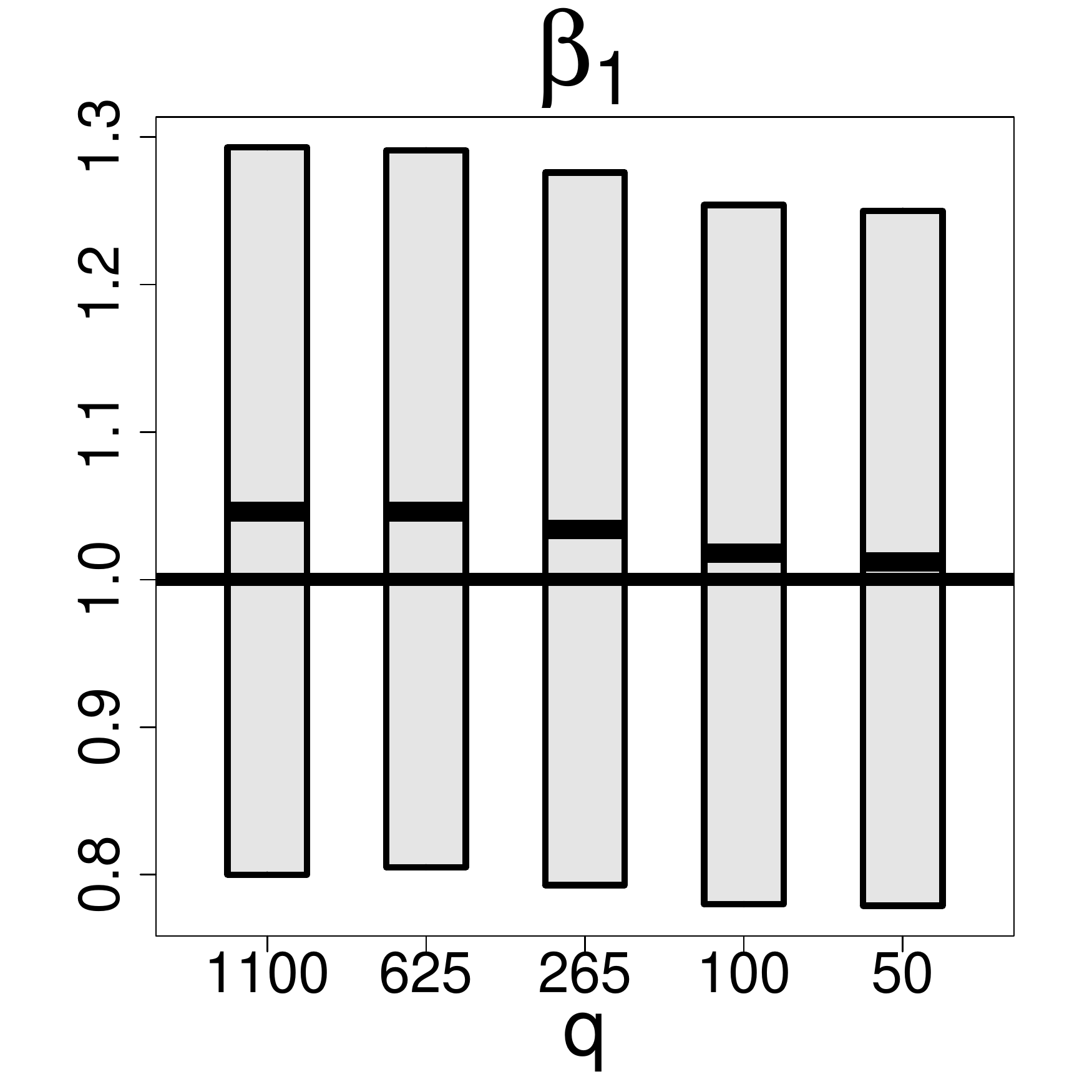} & \includegraphics[scale=.2]{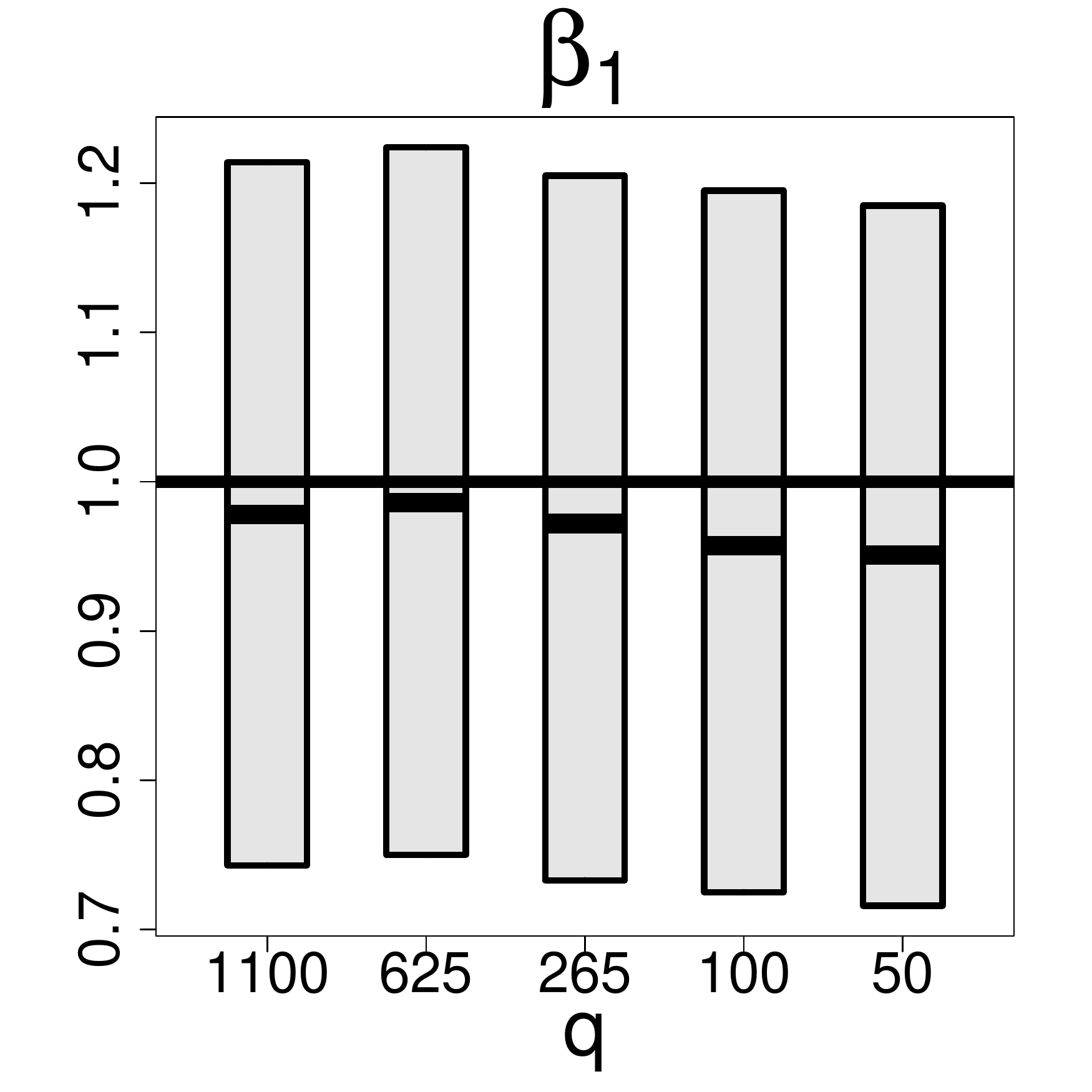}\\
 \includegraphics[scale=.2]{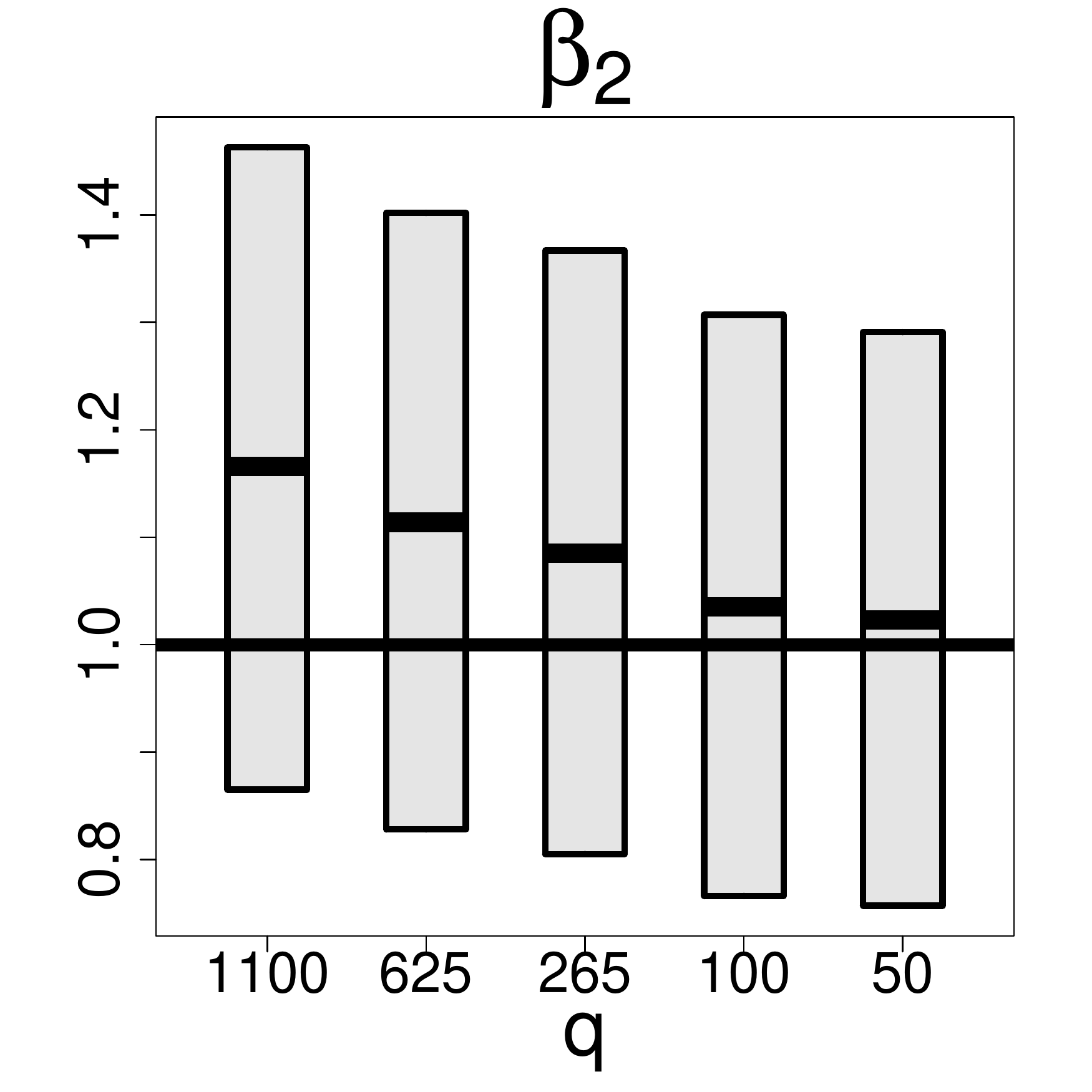} & \includegraphics[scale=.2]{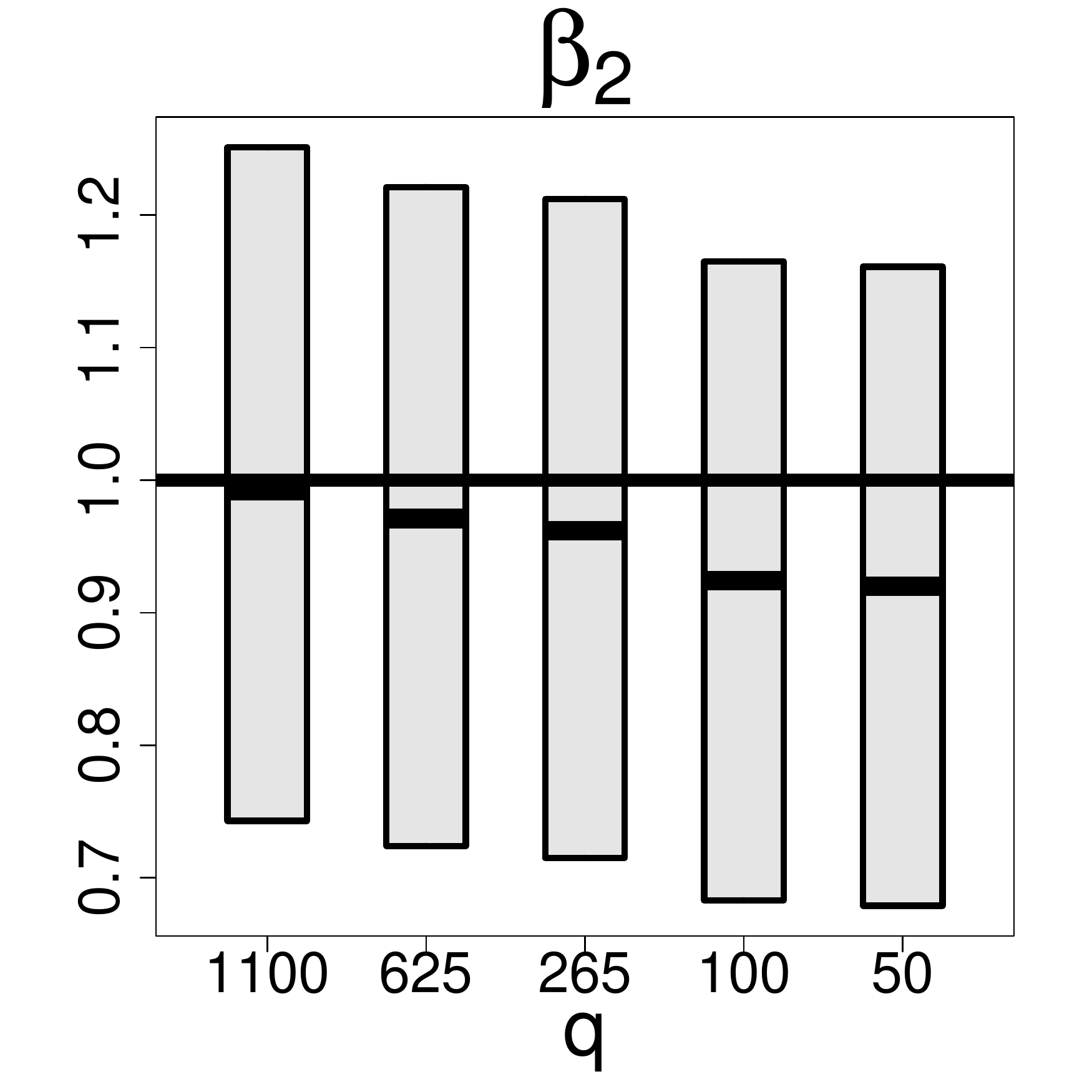} & \includegraphics[scale=.2]{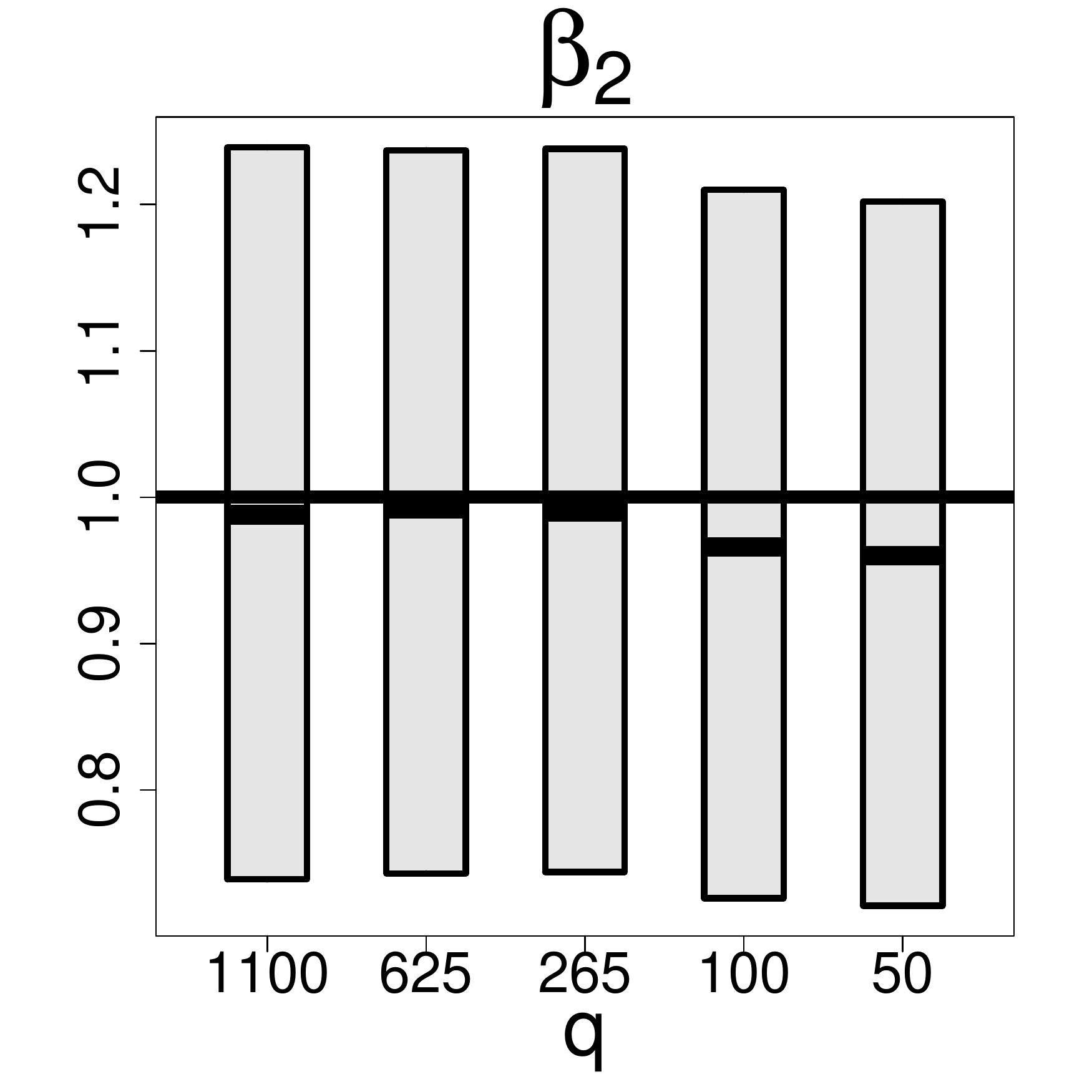} & \includegraphics[scale=.2]{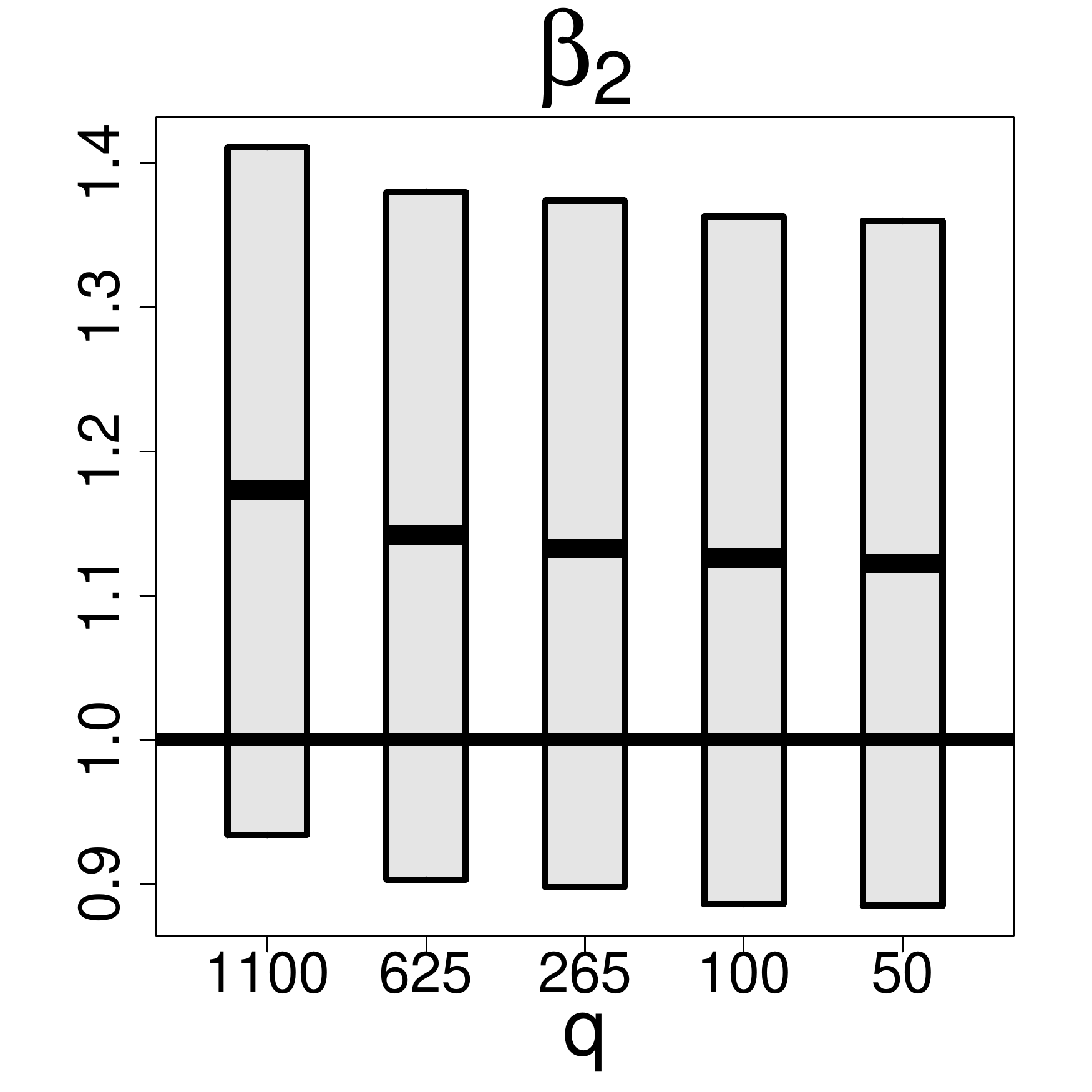}
\end{tabular}
   \caption{\label{heurbox}Box plots illustrating inference for $\bs{\beta}$ for simulated binary data with various values for $\tau$ and various degrees of sparsity.}
\end{figure}

The results of the study are shown in Table~\ref{heur} and Figure~\ref{heurbox}. Evidently, a very dramatic reduction in dimension comes at little cost to regression inference. In fact, since $\beta_1$ are $\beta_2$ are negatively correlated in this scenario, greater dimension reduction resulted in a better point estimate for $\bs{\beta}$ in many cases. And even using a mere 50 eigenvectors did not relocate any confidence interval enough to miss the true value of the parameter. Hence, 50--100 eigenvectors should suffice for most analyses, which removes evaluation of the above mentioned quadratic form as the chief computational burden in fitting the model. Should one wish to be more conservative, we recommend the use of all eigenvectors corresponding to standardized eigenvalues greater than 0.7. For a square lattice this means using approximately 10\% of the eigenvectors.

\begin{table}
\caption{\label{heur}The effect of various levels of sparsity, for simulated binary data with $\tau$ equal to 0.5, 1, 2, and 4. Running times are given in minutes.}
   \centering
   \setlength{\extrarowheight}{1.5ex}
   \begin{tabular}{|cccccccccc|}\hline
   $\tau$ & Dim & $\hat{\beta}_1$ & CI($\beta_1$) & $\hat{\beta}_2$ & CI($\beta_2$) & $\hat{\tau}$ & CI($\tau$) & $\Vert\bs{p}-\hat{\bs{p}}\Vert$ & Time\\\hline\hline
   \multirow{5}{*}{0.5} & 1100 & 0.862 & (0.581, 1.147) & 1.166 & (0.865, 1.463) & 0.665 & (0.439, 0.912) & 6.379 & 5,159\\
   & 625 & 0.869 & (0.592, 1.147) & 1.114 & (0.828, 1.402) & 0.617 & (0.372, 0.915) & 6.521 & 1,836\\
   & 265 & 0.857 & (0.588, 1.127) & 1.085 & (0.805, 1.367) & 0.512 & (0.311, 0.720) & 6.877 & 597\\
   & 100 & 0.852 & (0.588, 1.117) & 1.035 & (0.766, 1.307) & 0.464 & (0.253, 0.676) & 7.424 & 260\\
   & 50 & 0.832 & (0.571, 1.092) & 1.023 & (0.757, 1.291) & 0.410 & (0.206, 0.621) & 7.853 & 113\\\hline
   \multirow{5}{*}{1} & 1100 & 1.125 & (0.870, 1.380) & 0.992 & (0.743, 1.251) & 0.930 & (0.500, 1.319) & 5.205 & 6,067\\
   & 625 & 1.097 & (0.847, 1.344) & 0.971 & (0.724, 1.221) & 1.056 & (0.585, 1.687) & 5.294 & 1,666\\
   & 265 & 1.086 & (0.840, 1.337) & 0.962 & (0.715, 1.212) & 0.793 & (0.404, 1.198) & 5.418 & 579\\
   & 100 & 1.050 & (0.809, 1.293) & 0.924 & (0.683, 1.165) & 1.141 & (0.533, 1.865) & 5.787 & 169\\
   & 50 & 1.037 & (0.796, 1.275) & 0.920 & (0.679, 1.161) & 1.031 & (0.435, 1.726) & 6.005 & 109\\\hline
   \multirow{5}{*}{2} & 1100 & 1.046 & (0.800, 1.293) & 0.988 & (0.739, 1.239) & 2.234 & (0.676, 3.789) & 4.134 & 4,469\\
   & 625 & 1.046 & (0.805, 1.291) & 0.992 & (0.743, 1.237) & 1.701 & (0.856, 2.715) & 4.098 & 1,926\\
   & 265 & 1.034 & (0.793, 1.276) & 0.990 & (0.744, 1.238) & 1.507 & (0.548, 2.723) & 4.116 & 552\\
   & 100 & 1.018 & (0.780, 1.254) & 0.966 & (0.726, 1.210) & 2.031 & (0.616, 4.163) & 4.230 & 184\\
   & 50 & 1.012 & (0.779, 1.250) & 0.960 & (0.721, 1.202) & 2.478 & (0.465, 5.205) & 4.392 & 131\\\hline
   \multirow{5}{*}{4} & 1100 & 0.978 & (0.743, 1.214) & 1.173 & (0.934, 1.411) & 4.530 & (1.192, 6.581) & 3.238 & 4,698\\
   & 625 & 0.986 & (0.750, 1.224) & 1.142 & (0.903, 1.380) & 3.470 & (1.246, 7.501) & 3.417 & 2,316\\
   & 265 & 0.972 & (0.733, 1.205) & 1.133 & (0.898, 1.374) & 2.839 & (1.103, 5.186) & 3.506 & 546\\
   & 100 & 0.957 & (0.725, 1.195) & 1.126 & (0.886, 1.363) & 2.639 & (0.935, 4.939) & 3.567 & 249\\
   & 50 & 0.951 & (0.716, 1.185) & 1.122 & (0.885, 1.360) & 2.296 & (0.697, 4.580) & 3.652 & 155\\
   \hline
   \end{tabular}
\end{table}

\section{Application to US Infant Mortality Data}
\label{real}

\begin{figure}
\centering
 \includegraphics[scale=.7]{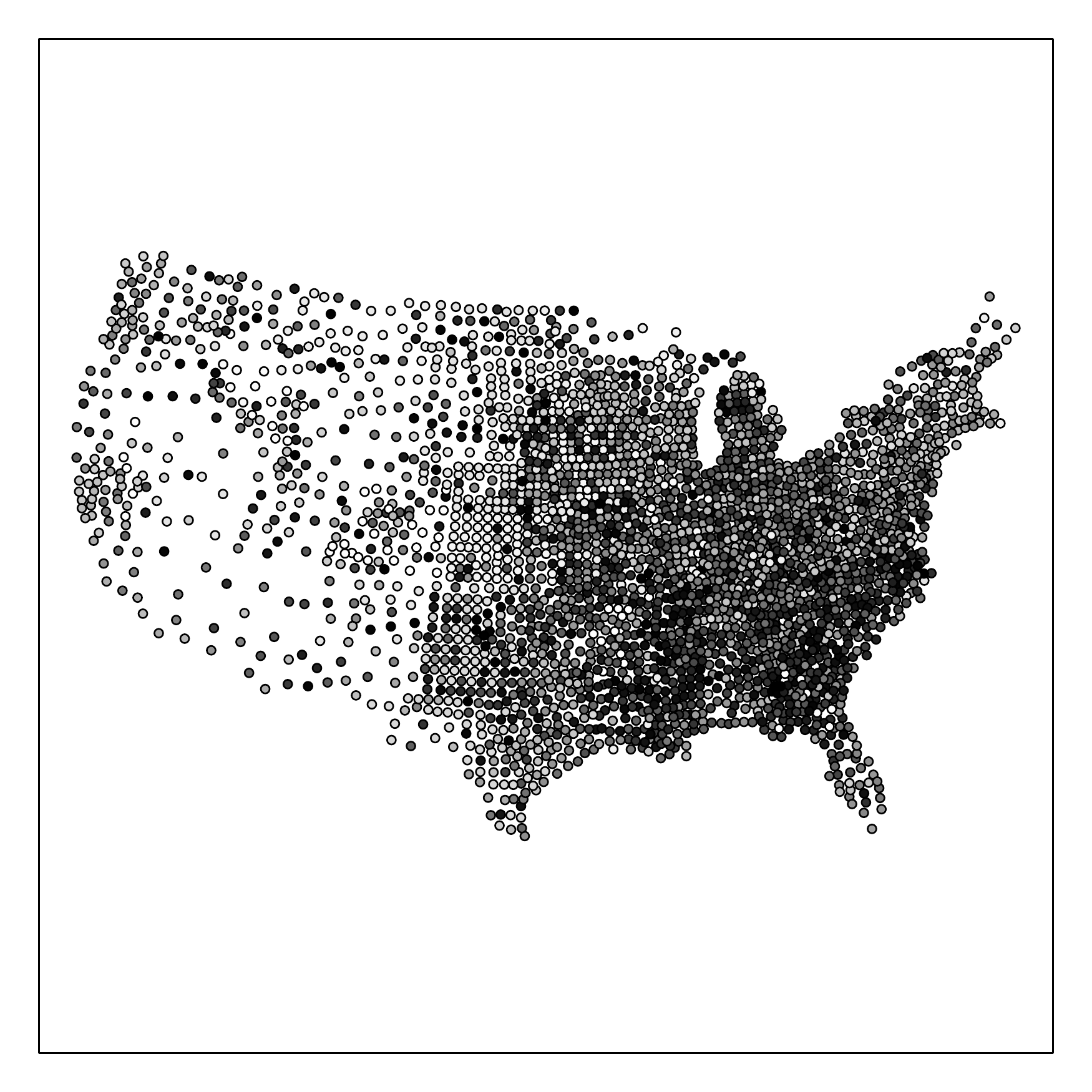}
   \caption{\label{usinfant}A plot of infant mortality rates for 3,071 counties of the contiguous United States. The shaded circle for a given county represents that county's ratio of deaths to births, where births and deaths were averaged over the three-year period 2002--2004. A darker circle indicates a higher rate.}
\end{figure}

The plot in Figure~\ref{usinfant} shows infant mortality data for 3,071 US counties. Each shaded circle represents a ratio of deaths to births, i.e., an empirical infant mortality rate, for a given county. The data were obtained from the 2008 Area Resource File (ARF), a county-level database maintained by the Bureau of Health Professions, Health Resources and Services Administration, US Department of Health and Human Services. Specifically, three variables were extracted from the ARF: the three-year (2002--2004) average number of infant deaths before the first birthday, the three-year average number of live births, and the three-year average number of low birth weight infants.

To these data we fit our sparse Poisson SGLMM with
\begin{align}
\label{infantmod}
\e(\textsc{deaths}_i\,|\,\bs{\beta},\bs{\delta}_S) &= \textsc{births}_i\,e^{\beta_0+\beta_1\textsc{low}_i+\beta_2\textsc{black}_i+\beta_3\textsc{hisp}_i+\beta_4\textsc{gini}_i+\beta_5\textsc{aff}_i+\beta_6\textsc{stab}_i+\mbf{M}_i\bs{\delta}_S},
\end{align}
where \textsc{deaths} is the number of infant deaths; \textsc{births} is the number of live births; \textsc{low} is the rate of low birth weight; \textsc{black} is the percentage of black residents (according to the 2000 US Census); \textsc{hisp} is the percentage of hispanic residents (2000 US Census); \textsc{gini} is the Gini coefficient, a measure of income inequality \citep{Gini1921Measurement-of-}; \textsc{aff} is a composite score of social affluence \citep{Yang:2009p918}; and \textsc{stab} is residential stability, an average $z$-score of two variables from the 2000 US Census. We chose $\dim(\bs{\delta}_S)=100$. Simulating a sample path of length 2,000,000 yielded Monte Carlo standard errors smaller than 0.005 for the regression coefficients, smaller than 0.1 for $\tau$, and on the order of 0.001 for the random effects. The analysis took just 3.5 hours. Fitting the RHZ model to these data took approximately fourteen days, partly because storing the sample paths for the 3,064 random effects required the use of a 46 GB file-backed matrix \citep{bigmem}. Had we been able to store all results in RAM, fitting the RHZ model would have taken approximately four days.

The results are given in Table~\ref{tabinfant}. We see that all predictors were found significant. The posterior mean for $\tau$ is much smaller than the prior mean of 1,000, which contraindicates a nonspatial analysis of these data. Moreover, five eigenvectors were found to be significant predictors.

\begin{table}
\caption{\label{tabinfant}The results of fitting model (\ref{infantmod}) to the infant mortality data.}
\centering
\begin{tabular}{|cccc|}\hline
Predictor & Parameter & Estimate & CI \\\hline\hline
Intercept & $\beta_0$ & -5.452 & (-5.638, -5.268)\\
\textsc{low birth weight} & $\beta_1$ & 8.719 & (7.475, 9.971) \\
\textsc{black} & $\beta_2$ & 0.00428 & (0.00295, 0.00559) \\
\textsc{hispanic} & $\beta_3$ & -0.00405 & (-0.00515, -0.00291) \\
\textsc{gini} & $\beta_4$ & -0.495 & (-0.930, -0.0610) \\
\textsc{affluence} & $\beta_5$ & -0.0774 & (-0.0896, -0.0654) \\
\textsc{stability} & $\beta_6$ & -0.0292 & (-0.0439, -0.0144) \\
-- & $\tau$ & 12.016 & (6.088, 19.116) \\
\hline
\end{tabular}
\end{table}

\section{Discussion}

In this paper we developed a framework for reparameterization of the areal SGLMM. By accounting appropriately for the covariates of interest, our approach alleviates the spatial confounding that plagues the traditional SGLMM. And, by accounting for the underlying graph, our model (1) permits patterns of positive spatial dependence only and (2) allows for dramatic reduction in the dimension of the random effects.

We studied the performance of our approach for three of the most commonly used first-stage models: Bernoulli, Poisson, and Gaussian. In all three cases our approach resulted in better regression inference and in considerable gains in the computational efficiency of the MCMC algorithms used for inference. A thorough followup simulation study indicates that a small, fixed number of random effects is sufficient for all datasets. More conservatively, our sparse model appears to require no more than 0.1$n$ random effects to perform well with respect to both regression and prediction. The resulting gain in computational efficiency will permit the relatively rapid analysis of datasets that were once considered too large for the areal SGLMM.

\section*{Acknowledgements}

We thank Tse-Chuan Yang for providing the infant mortality dataset and for helpful discussions.

\bibliography{sglmm}
\bibliographystyle{chicago}

\end{document}